\documentclass[aps,prd,twocolumn,amsmath,amssymb,nofootinbib]{revtex4}

\usepackage[dvips]{epsfig}
\usepackage{amssymb}
\usepackage{graphicx}
\usepackage{graphics}
\usepackage{amsfonts}
\usepackage{amsmath}
\usepackage{color}
\usepackage{latexsym}
\usepackage{psfrag}

\definecolor{Blue}{rgb}{0.3,0.3,0.9}
\definecolor{Red}{rgb}{1.0,0.0,0.0}
\definecolor{Green}{rgb}{0,0.4,0}
\definecolor{Violet}{rgb}{0.4,0.0,0.6}
\definecolor{Cyan}{rgb}{0.0,0.4,0.6}
\definecolor{Orange}{rgb}{1.0,0.4,0.0}

\newcommand{\ket}[1] {\mbox{$ \vert #1 \rangle $}}

\newcommand{\bra}[1] {\mbox{$ \langle #1 \vert $}}

\newcommand{\inv}[1]{\frac{1}{#1}}

\newcounter{subequation}[equation] \makeatletter
\expandafter\let\expandafter\reset@font\csname
reset@font\endcsname
\newenvironment{subeqnarray}
  {\arraycolsep1pt
    \def\@eqnnum\stepcounter##1{\stepcounter{subequation}{\reset@font\rm
      (\theequation\alph{subequation})}}\eqnarray}
  {\endeqnarray\stepcounter{equation}}
\makeatother

\newcommand{\ba}{\begin{eqnarray}}
\newcommand{\ea}{\end{eqnarray}}
\newcommand{\sba}{\begin{subeqnarray}}
\newcommand{\sea}{\end{subeqnarray}}
\newcommand{\bwt}{\begin{widetext}}
\newcommand{\ewt}{\end{widetext}}

\def\t{\tau}

\def\ga{\gamma}

\def\la{\lambda}

\def\al{\alpha}

\def\e{\epsilon}

\def\di{\partial}

\def\didiv{\raise 0.1mm \hbox{$\stackrel{\leftrightarrow}{\di_V}$}}
\newcommand{\didi}[1]{\raise 0.1mm \hbox{$\stackrel{\leftrightarrow}{\di_{#1}}$}}

\def\re{\mbox{Re}}
\def\im{\mbox{Im}}

\def\1{{\mathbf 1}}

\def\bitem{\begin{itemize}}
\def\eitem{\end{itemize}}
\def\bes{\begin{description}}
\def\es{\end{description}}
\newcommand{\be}{\begin{equation}}
\newcommand{\ee}{\end{equation}}

\def\inte{\int_{-\infty}^{+\infty}}

\newcommand\Ie[1]{\inte \!\! d #1 \;}

\overfullrule=0pt \def\sqr#1#2{{\vcenter{\vbox{\hrule height.#2pt
          \hbox{\vrule width.#2pt height#1pt \kern#1pt
           \vrule width.#2pt}
           \hrule height.#2pt}}}}
\newcommand\lrpartial[1]{\mathrel{\partial_{#1}\kern-1em\raise1.75ex\hbox{$\leftrightarrow$}}}

\begin{document}

 \title{Why does the Unruh effect rely on Lorentz invariance, \\ 
while Hawking radiation does not ?} 
 \author{David Campo}
 \email[]{dcampo@astro.physik.uni-goettingen.de}
 \affiliation{Georg-August-Universit\"{a}t,
 Institut f\"{u}r Astrophysik,
 Friedrich-Hund-Platz 1, 37077 G\"{o}ttingen, Germany}
\author{Nathaniel Obadia}
\email[]{nathaniel.obadia@ens-lyon.fr}
\affiliation{Centre de Recherche Astrophysique de Lyon, CNRS, UMR 5574; \'Ecole Normale
Sup\'erieure de Lyon, 46 all\'ee d'Italie, F-69364 Lyon cedex 07, France;
Universit{\'e} de Lyon, Lyon, F-69003, France; Universit\'e Lyon 1,
Villeurbanne, F-69622, France}
 \begin{abstract}
We show that without Lorentz invariance, 
the Unruh effect does not exist. 
We use modified dispersion relations and describe in turn: 
the non-thermal nature of the vacuum (defined in the preferred frame) 
restricted to the Rindler wedge, the loss of the KMS property of the Wigthman function,
the transition amplitudes and transition rates of a 
uniformaly accelerated detector.
This situation seems to contrast with the Hawking radiation of acoustic black holes,
which under certain assumptions has been shown 
to be robust to a breaking of Lorentz symmetry.
We explain this discrepancy.
 \end{abstract}
\maketitle

It is well understood that Unruh effect and Hawking radiation are
physically distinct phenomena but with a common root, 
if not a physical at least a mathematical one \cite{redbookWald}.
Yet, the reason of this analogy is not fully understood.

The situation seems in fact paradoxical. 
On one hand, we are going to show that 
the Unruh effect is inseparable from Lorentz symmetry:
without the latter, the former does not exist.
On the other hand, the study of black holes analogues in supersonic fluids and 
Bose-Einstein condensates,
the so-called acoustic black holes or dumb holes, has shown (under certain 
assumptions) that the defining properties of Hawking radiation (in that case 
a stationary and thermal flux of phonons escaping from 
the sonic horizon to infinity) are robust 
to a breaking of Lorentz symmetry, see \cite{acoustic} for a review. 
So how can they be related ? 
A reconciliation of these two results seems necessary in view of the
analogy mentioned above.

The fundamental role played by Lorentz symmetry in the existence of the Unruh effect 
is particularly clear from the algebraic proof of this one \cite{BandW,Sewell}. 
This proof establishes, at the level of the operator algebra, that 
the Minkowski vacuum restricted to a Rindler wedge
is a thermal state with respect to the boost parameter.
This theorem is proved in the framework of 
axiomatic field theories, which makes the instrumental
role of Lorentz symmetry abundantly clear.
Yet one can wonder whether, in the absence of the invariance under boosts, 
the Unruh effect exists in a more operational sense (outside transients of course).
As we will show, the answer is negative.

We conducted a complete analysis of the Unruh effect 
without Lorentz invariance. We examine the mapping between Minkowski and Rindler Fock 
spaces (Unruh modes and Bogoliubov transformations), the Wightman function, and the 
transition amplitudes and transition rates of a detector coupled to the field. 
Lorentz symmetry considerably constrains the structure of field theories
and renders all these descriptions of the effect equivalent.
The unifying role of Lorentz symmetry and the instrumental role of
the stable vacuum are recalled in sec. \ref{sec:withLI}.
Without Lorentz symmetry, each quantity answers to 
a different question as will be explained in sec. \ref{sec:wihtoutLI}.
Finally we will compare Unruh and Hawking effects in section \ref{sec:comparison}.
The reader will find the main results summarized and discussed in the
subsection $E$ of each section.

\section{Unruh effect with Lorentz invariance}
\label{sec:withLI}

We set up our notations and review the Lorentz invariant case.
The contend of this section is not new but our presentation, which lays 
emphasis on the role of Lorentz symmetry, may be original, see in particular 
sec. \ref{sec:double role}.
We refer to the review articles \cite{GrandeOeuvre,Matsas} for more details.

We consider the simple case of a free massless scalar field $\varphi(t,{\bf x})$
in Minkowski space-time. In inertial frames equiped with a global 
and cartesian coordinate system, i.e.   
$ds^2 = - dt^2 + \delta_{ij} dx^i dx^j$ with units $c=1$,
the wave equation is 
\ba \label{eom Mink}
  \left( \partial_t^2 - \delta^{ij}\partial_i\partial_j \right) \varphi = 0  \, .
\ea
The field is assumed to be neutral so its creation and annihilation parts
are conjugate from one another
\ba \label{Phi}
   \varphi(t,{\bf x}) = 
  \int\!\!d^3k \, \left( a_{\bf k} \varphi_{\bf k}(t,{\bf x}) + 
   a_{\bf k}^{\dagger} \varphi_{\bf k}^*(t,{\bf x}) \right) \, ,
\ea
where  
\ba \label{Minkmodes}
  \varphi_{\bf k}(t,{\bf x}) = 
  \frac{e^{-i\omega_k t + i {\bf k} {\bf x}}}{\sqrt{(2\pi)^3 \, 2\omega_k}} 
\ea
are the plane wave solutions of (\ref{eom Mink}) with positive frequency
w.r.t the inertial time coordinates, i.e. 
$i\partial_t \varphi_{\bf k} = \omega_k \varphi_{\bf k} \, , \, \omega_k > 0$.
For massless fields the dispersion relation is of course 
$\omega_k = k \equiv \vert {\bf k} \vert$.
The state of the field is the Lorentz invariant ground state
defined by $a_{\bf k}\ket{0_M}=0$ for all $\bf k$.

A two-level point detector is coupled to the field as described by the 
interaction Hamiltonian
\ba \label{Hint}
  H_{int} = g \, 
   \left( e^{iE\tau} \ket{+} \bra{-} 
  +  e^{-iE\tau} \ket{-} \bra{+} \right) \, \varphi[x^a(\tau)] \, .
\ea
$g$ is a dimensionless coupling constant.
$\ket{\mp}$ are respectively the ground and excited energy eigenstates.
They are separated by the energy gap $E > 0$ in the detector's rest frame. 
$\tau$ is the detector's proper time and $x^a(\tau)$ the detector's trajectory.
The latter will be constrained to be linear, and we can choose the 
coordinates such that the detector moves along the $z$-axis.
Inertial (In) timelike trajectories with velocity $\beta$ are therefore parametrized by
\ba \label{Intraj}
   \mbox{Inertial} &:& [\ga\t, \, 0, \, 0,\, \gamma \beta\t] \ ,
\ea
where $\gamma = (1-\beta^2)^{-1/2}$ is the Lorentz factor.
Uniformly Accelerated (UA) detectors with acceleration
$a=\left(\eta_{ab} \frac{d^2x^a}{d\tau^2}  \frac{d^2x^b}{d\tau^2}\right)^{1/2}$ 
follow the trajectory
\ba \label{UAtraj}
   \mbox{UA} &:& [ a^{-1}\sinh(a\t) ,\, 0,\, 0, \, \pm \, a^{-1} \cosh(a\t)] \, .
\ea
The corresponding Lorentz factor is
\ba \label{Gamma}
  \Gamma(\tau) \equiv \frac{d t}{d \tau} = \cosh(a\tau) \, .
\ea

The literature on the Unruh effect is sometimes confused on 
the definition of the effect.
There are two different aspects to it. The first aspect is 
the thermal nature of the Minkowski vacuum for observables 
with support in a Rindler wedge.
The R(ight) and L(eft) Rindler wedges of Minkowski space are the open sets
$\left\{ (t,x,y,z) / z \lessgtr 0 ,\, \vert t \vert < \pm z  \right\}$.
They are both static, globally hyperbolic space-times, so that
a consistent quantum theory can be defined on them \cite{Fulling}.
This theory is not equivalent to
the usual quantization in the full Minkowski space-time.
Rather, the Minkowski vacuum corresponds to a thermal state of Rindler quanta 
at "temperature $1/2\pi$" (in arbitrary units).
The second aspect is the operational meaning of the previous result
in terms of the response of a UA detector \cite{Unruh}.
(As we can see from (\ref{UAtraj}), 
a UA trajectory is confined to a Rindler wedge.)
To wit, the $S$-matrix elements and transition rates of the detector
verify detailed balance.

The thermal nature of the Minkowski vacuum restricted to 
a Rindler wedge is recalled in section \ref{sec:stateLI}
and the Wightman function is described in \ref{sec:WLI}.
We then consider the dynamics of the detector 
introduced at eq. (\ref{Hint}).
We will see respectively in sec. \ref{sec:ampLI} and \ref{sec:ratesLI} 
that the Bogoliubov coefficients of section \ref{sec:stateLI}
are proportional to the $S$-matrix elements 
of the processes $\ket{\mp} \to \ket{\pm} + {\bf k}$, 
and that the Wightman function
transmits its properties to the transition rates
which are essentially the Fourier transform of the former.
We will insist on how the results are related by Lorentz symmetry.
We therefore do not choose to define the Unruh effect by one or the other
aspect because by Lorentz invariance they are equivalent.

\subsection{The Minkowski state for observables in a Rindler wedge}
\label{sec:stateLI}

The proof begins with the definition of the field theory in the 
static, globally hyperbolic Rindler wedges.
The metric in the $R$ wedge can be brought in the static form by the 
following change of coordinates
\ba \label{coordR}
   t = \kappa^{-1} e^{\kappa \zeta} \sinh(\kappa\eta) 
   \,  , \quad 
   z = \kappa^{-1} e^{\kappa \zeta} \cosh(\kappa \eta) 
\ea
where $\kappa$ is an arbitrary energy scale.
We choose units $\kappa = 1$. 
The line element is $ds^2 = e^{2\zeta} (- d \eta^2 + d \zeta^2) + dx^2 + dy^2$.
The timelike Killing vector $\partial_\eta$ 
corresponds in the $R$ wedge to 
the generator $z \di_t + t \di_z$ of boosts in the $z$-direction.
The future and past horizons are respectively 
\ba \label{hor}
  {\cal H}^{\pm} = \left\{ \eta \to \pm \infty , \, \zeta \to - \infty ,\, 
 \eta \pm \zeta \,\,\, {\rm finite} \right\} \, .
\ea
The null coordinates $u = \eta - \zeta$ and $v = \eta + \zeta$ will be usefull.
Letting ${\bf x}_\bot=(x,y)$, the field equation (\ref{eom Mink}) in these coordinates is
\ba \label{waveeqR}
  \left(\di_\eta^2-\di_\zeta^2- e^{2\zeta}\left(\di_x^2+\di_y^2\right)\right) 
  \varphi(\eta,\zeta, {\bf x}_\bot) &=& 0  \, .
\ea
Since the metric is static, the solutions of (\ref{waveeqR}) can be classified
according to the eigenfunction of the timelike Killing vector field 
$i\di_\eta = \la$ where $\lambda > 0$, given by
\ba \label{phiR}
   \varphi^R_{\la,k_\bot}(\eta,\zeta,{\bf x}_\bot)
   = \frac{\sqrt{\sinh(\pi\la)}}{2\pi^2} 
   K_{i\la}(k_\bot e^{\zeta}) \, 
   e^{-i(\la \eta -{\bf k}_\bot {\bf x}_\bot)}  \, . \quad 
\ea
Similarly in the L quadrant we can introduce the coordinates $(\bar \eta, x, y, \bar \zeta)$
defined by
\ba \label{coordL}
   t= - \, e^{\bar \zeta} \sinh(\bar \eta) \,  , \quad 
   z= - \, e^{\bar \zeta} \cosh(\bar \eta) 
\ea
We chose $dt/d\bar \eta < 0$ because the boost Killing vector field
$t\partial_z + z\partial_t$ is timelike and past-directed in that wedge.
This convention implies that the modes defined by
\ba \label{phiL}
   \varphi^L_{\la,k_\bot}(\bar \eta, \bar \zeta,{\bf x}_\bot) = 
    \varphi^{R \, *}_{\la,k_\bot}(\bar \eta,\bar \zeta,{\bf x}_\bot) \, , \quad \lambda > 0 \, .
\ea
are positive frequency Rindler modes.
This double familly of modes form a complete orthonormal basis on which
the field can be decomposed
\ba\label{PhiRindler}
  \varphi(\t,\zeta,{\bf x}_\bot)
   = \int_{0}^{\infty}\!\!d\lambda \int\!\!d^2k_\bot  
     \left( a^R_{\la,k_\bot}\varphi^R_{\la,k_\bot} +  
    a^L_{\la,k_\bot}\varphi^{L \, *}_{\la,k_\bot}  + h.c. \right)
\nonumber \\
\ea
We designate by $\ket{0_{R,L}}$ the vacua in the 
$R$ and $L$ wedges respectively.

To an observer living in the $R$-wedge, that is an 
observer measuring observables $\hat {\cal O}_R \otimes {\bf 1}_L$, 
the Minkowski vacuum appears to be the mixed state $\rho$ defined by
\ba 
  {\rm Tr}\left( \rho \, \hat {\cal O}_R \right)
  \equiv {\rm Tr}\left( \ket{0_M} \bra{0_M} \hat {\cal O}_R \otimes {\bf 1}_L \right)
\ea
for every observable, that is
\ba \label{rho}
  \rho = {\rm Tr}_L\left(  \ket{0_M} \bra{0_M} \right)
\ea
where the partial trace ${\rm Tr}_L$ is over the Hilbert space of the 
theory defined in the $L$-wedge.
If the field theory is free, it is possible to calculate $\rho$ explicitely.
To this end, one first establishes the unitary map between $\ket{0_M}$ and 
the Rindler Fock states, or equivalently, since the theory is free,
a unitary transformation between the positive frequency Minkowski modes (\ref{Minkmodes})
and the Rindler modes (\ref{phiR}) and (\ref{phiL}).
From now on we do not write the subscript ${\bf k}_\bot$ anymore since it 
is obvious that the map does not mix different transverse wavevectors.

This map is found by appealing for the stability of the Minkowski vacuum and
the concomitant analytic properties of the Minkowski modes. 
Namely, the stability of the vacuum in any inertial frame is tantamount to the 
analyticity and boundedness of the positive frequency Minkowski modes (\ref{Minkmodes}) 
in the domain 
${\cal T} = \left\{ x+iy | y^0 < 0 \, , \vert y^0 \vert \geq \vert {\bf y} \vert \right\}$. 
Since any linear combination  
of positive frequency Minkowski modes is also analytic and bounded in ${\cal T}$,
the idea is to define a new familly of modes $\varphi^U_{\Omega}$,
defined in the complete Minkowski space,
such that i) they are eigenmodes of the boost Killing  vector
$z\partial_t + t \partial_z$, hence of the form (no Rindler-frequency mixing)
\sba \label{phiU}
  \Omega > 0 \, , \quad 
  \varphi^U_\Omega &=& \alpha_\Omega \varphi_\Omega^R + \beta_\Omega \varphi_\Omega^{L \, *}
\\
  \Omega < 0 \, , \quad 
  \varphi^U_\Omega &=& \alpha_{\vert \Omega \vert} \varphi_{\vert \Omega \vert}^L + 
   \beta_{\vert \Omega \vert} \varphi_{\vert \Omega \vert}^{R \, *} \, , 
\sea
ii) admit the decomposition into positive frequency Minkowski modes
\sba \label{CL mink}
  \Omega > 0 \, , \quad  \varphi^U_{\Omega} &=& \int_{-\infty}^{+\infty}\!\!dk_z \, 
   A_{\Omega, k_z} \, \varphi_{k_z}^M \, , 
\\
  \Omega < 0 \, , \quad  \varphi^U_{\Omega} &=& \int_{-\infty}^{+\infty}\!\!dk_z \, 
   B_{\Omega, k_z} \, \varphi_{k_z}^{M \,*} \, ,
\sea
iii) and are orthonormal and complete (w.r.t. the Klein-Gordon product).
This basis of solutions of the wave equation, called the Unruh modes, 
characterizes the Minkowski vacuum and contains within its definition 
(\ref{phiU}) the mapping we are looking for \cite{Unruh}.

For the function $\varphi^U_\Omega$ introduced in eqs. (\ref{phiU}) to be
defined on the entire Minkowski space, we need to extend the 
definition of the Rindler modes $\varphi^{R}$ 
to the $L$-wedge. From now on $\varphi^R_\lambda$ denotes the function
equal to the r.h.s. of (\ref{phiR}) in the $R$-wedge, and equal to zero in the $L$-wedge.
We define similarly $\varphi^L_\lambda$ the extension of the $L$-Rindler modes
(\ref{phiL}) to the $R$-wedge.

The task to find the Bogoliubov coefficients $\alpha_\Omega$ and $\beta_{\Omega}$
such that $\varphi^U_\Omega$ as defined by eqs. (\ref{phiU}) verifies condition ii)
is simplified by the following theorem \cite{redbookWald}: 
any solution of the Klein-Gordon
equation in dimensions more than $2$ is characterized by its 
restriction to either ${\cal H}^+$ or ${\cal H}^-$.
Written in terms of the null coordinates $U = t-z$ and $V = t + z$, 
the positive frequency Minkowski modes 
\ba
  \varphi_{k_z}^M \propto e^{-i (\omega_k - k_z) V/2 - i  (\omega_k + k_z) U/2 }
\ea
are analytic and bounded functions for the complex values of $U$ and $V$
such that $\im(V) < 0$ and $\im(U) < 0$ because 
$\omega_k^2 = k_z^2 + {\bf k}_\bot^2 \geq k_z^2$.
We therefore require that the restriction of the linear combination (\ref{phiU})
on, say the future horizon shares the same property. 
The asymptotic form of (\ref{phiR}) near ${\cal H}^+$ is the sum of two terms
\ba \label{asymptphiR}
  \varphi^R_{\lambda}(\tau,\zeta) \sim N_{\lambda} \left( 
  \frac{(k_\bot/2)^{i\lambda}}{\Gamma(1+i\lambda)} e^{-i\lambda u} - 
  \frac{(k_\bot/2)^{-i\lambda}}{\Gamma(1-i\lambda)} e^{-i\lambda v} \right) \quad
\ea  
We do not need the expression of $N_\lambda$ for the moment.
On ${\cal H}^+$, $u=\infty$ and $v$ is finite, so the first term is singular.
When the modes are superposed to form wave-packets, 
the term which rapidly oscillates as the wave-packet nears the horizon does not
contribute. The reason for considering wavepackets is that fields are 
operator valued distributions, and therefore must be smeared. (We did not bring up
this issue before because it is the only place in this section where 
it is mandatory to use wavepackets instead of modes.)
So we retain only the second term.
Since the null coordinates are related by $V \vert_R = e^v$ 
and $V\vert_L = - e^{\bar v}$, the r.h.s. of (\ref{phiU}) is 
\ba \label{asymp phiU}
  \Omega > 0 \, , \, 
   \varphi^U_\Omega &\sim& \alpha_\Omega \vert V\vert^{-i\Omega} \left( \theta(V) +  
   \theta(-V) \frac{\beta_\Omega}{\alpha_\Omega}  \right) \, , 
\ea
and similarly for $\Omega < 0$.
The unique analytic continuation of $V^{-i\Omega}$ which is bounded in $\im(V)< 0$
is $(V-i\epsilon)^{-i\Omega}$ where the branch cut of the logarithm 
extended to the complex plane is chosen along the negative real axis.
The relative weight is therefore
\ba \label{Bratio}
  \frac{\beta_\Omega}{\alpha_\Omega} = e^{-\pi \vert \Omega\vert } \, ,
\ea
and $\varphi^U_\Omega$ is normalised if
\ba \label{alpha}
  \vert \alpha_\Omega \vert^2 = \frac{1}{1-e^{-2\pi\vert \Omega\vert }} \, .
\ea
$\alpha_\Omega$ can be chosen real.

A direct calculation finally shows that the $\varphi^U_\Omega$ form a complete familly, so 
that the field can be represented as follows
\ba \label{phi Unruhmodes}
  \varphi = \int_{-\infty}^{\infty}\!\!d\Omega \int\!\!d^2k_\bot \, 
  \left( a_\Omega^U \varphi^U_\Omega + h.c. \right)
\ea
and the Minkowski vacuum is characterized by
\ba \label{charact vac}
  a_\Omega^U \ket{0_M} = 0 \, .
\ea
One can invert the relations (\ref{phiU}) to relate the creation and annihilation 
operators
\ba \label{Bogol operateurs}
  \Omega > 0 \, , \quad   
  a_\Omega^U &=& \alpha_\Omega a_{\Omega}^R - \beta_\Omega a_{\Omega}^{L \, \dagger} 
         \equiv {\cal U} a_{\Omega}^R {\cal U}^{-1} \, , 
\nonumber \\
  \Omega < 0 \, , \quad   
  a_\Omega^U &=& \alpha_{\vert \Omega \vert} a_{\vert \Omega \vert}^{L} 
  - \beta_{\vert \Omega \vert} a_{\vert \Omega \vert}^{R \, \dagger} 
         \equiv {\cal U} a_{\vert \Omega \vert}^L {\cal U}^{-1} \, , \quad
\ea
where ${\cal U}$ is the squeezing operator
\ba \label{map}
  {\cal U} = \prod_{\Omega} \exp\left\{ 
  {\rm argth}\left( \frac{\beta_\Omega}{\alpha_\Omega} \right) 
  \left( a_{\vert \Omega\vert }^{R\, \dagger} a_{\vert \Omega\vert }^{L\, \dagger} 
  - a_{\vert \Omega\vert }^{R} a_{\vert \Omega\vert }^{L} \right)  \right\}
\ea
This is the unitary map we where after. The Minkowski vacuum is indeed related to 
the Rindler Fock states by 
\ba \label{entangled}
  \ket{0_M} &=& {\cal U} \ket{0_R} \ket{0_L}
\nonumber \\
&=& \prod_{\Omega, {\bf k}_\bot}
  \sum_{n=0}^{\infty} e^{-n 2\pi \vert \Omega \vert} \ket{n_R}\ket{n_L} 
  \, .
\ea
In consequence, an observer living in say the $R$ wedge 
interpretes the Minkowski vacuum as a thermal bath of $R$-quanta
at the temperature $T= 1 / 2\pi$
\ba \label{rhoth}
   \rho = \prod_{\Omega, {\bf k}_\bot} e^{-n 2\pi \vert \Omega \vert } \ket{n_R} \bra{n_R}
\ea

We finally mention that 
the Bogoliubov coefficients have two different physical interpretations.  
The first is provided by (\ref{Bogol operateurs}):
they describe the mean values and correlations of Rindler quanta in the 
Minkowski vacuum
\ba \label{interp1}
  \bra{0_M} a_{\lambda}^{R\,\dagger}  a_{\lambda'}^{R} \ket{0_M} 
  = \delta(\lambda - \lambda')  \, \vert \beta_{\lambda} \vert^2 
\, , \nonumber \\
  \bra{0_M} a_{\lambda}^{R}  a_{\lambda'}^{L} \ket{0_M} 
  = \delta(\lambda - \lambda')  \, \alpha_{\lambda} \beta_{\lambda}  \, .
\ea
The second interpretation is given by their proportionality to the 
$S$-matrix elements, see eq. \ref{A proto alpha}.
These identities as well as the expressions of the coefficients $A$ and $B$ 
in eq. (\ref{CL mink}) are established in appendix \ref{app:Bogol}.

\subsection{Wightman function}
\label{sec:WLI}

The interpretation of $\ket{0_M}$ either as the ground state or a thermal state 
is reflected in the corresponding expressions of the two-point Wightman function.
For arbitrary events with inertial coordinates $y$ and $x+y$, 
\ba
  W(x) &=& \bra{0_M} \varphi(x+y) \varphi(y) \ket{0_M} 
\nonumber \\  
\label{Wint}
  &=& \frac{-i}{8\pi^2 \vert {\bf x} \vert}\int_0^\infty\!\!dk \, e^{-i\omega_k x^0}
  \left( e^{i k \vert {\bf x} \vert} 
  - e^{-i k \vert {\bf x} \vert}\right) e^{-\epsilon k}
\nonumber \\ 
\label{W}
  &=& -\frac{1}{4\pi^2} \frac{1}{(x^0 -i\epsilon)^2 -{\bf x}^2}
  \, .
\ea
The last expression exhibits the analyticity of $W$ 
in the lower half complex $x^0$-plane
and can be obtained from the integral representation by the introduction
of a regulator $e^{-\epsilon k}$.
This regulator implements the stability of the vacuum since, as we already saw,
the Minkowski modes (\ref{Minkmodes}) are
analytic and bounded in the complex space-time domain 
${\cal T} = \left\{  x+iy | \, y^0 < 0 \, , \vert y^0\vert \geq \vert {\bf y} \vert  \right\}$.
By Lorentz invariance, 
this in turn is equivalent to say that the one dimensional section of the 
Wightman function on the inertial straight lines is an analytic function
of the complexified affine parameter $\tau$ of the geodesics in the lower half plane 
$\im(\tau)< 0$.
The expression of the Wightman function, when both points 
belong to a common inertial trajectory (\ref{Intraj}),
\ba \label{WIn}
  {\cal W}_{In}(\tau) &\equiv& 
  \bra{0_M} \varphi[x_{In}(\tau+\tau')] \, \varphi[x_{In}(\tau')] \ket{0_M} 
  \nonumber \\ 
  &=& -\frac{1}{4\pi^2} \frac{1}{(\tau -i\epsilon)^2}
\ea
is indeed an analytic function in the lower half complex $\tau$-plane,
where $\tau^2 = (x^0)^2 - \vert {\bf x} \vert^2$.

The expression of the Wightman function 
when $x$ and $y$ are arbitrary points in the $R$ wedge 
is not illuminating. It becomes interesting 
only if the points are on an orbit of the generators of boosts such as the 
linearly UA trajectory (\ref{UAtraj}) 
\ba \label{WUA}
   {\cal W}_{ua}(\tau) &\equiv& 
   \bra{0_M} \varphi[x_{ua}(\tau+\tau')] \, \varphi[x_{ua}(\tau')] \ket{0_M} 
  \nonumber \\
   &=& -\frac{a^2}{16\pi^2} \, \frac{1}{\sinh^2\left( \frac{a\tau}{2} -i\epsilon \right)}
\ea
This function enjoys two important properties.
First, it depends only on the difference $\tau$ of the proper times, 
because the  vacuum is Lorentz invariant and a shift of $\tau$ corresponds to a boost.
Second, (\ref{WUA}) verifies the KMS condition
\ba \label{KMS}
  {\cal W}_{ua}\left(\tau + i\frac{2\pi}{a}\right)  = {\cal W}_{ua}(-\tau)
\ea
which is the definition of an equilibrium state of temperature $a/2\pi$.

Finally, the expression of the Wightman function on two points in 
opposite Rindler wedges, both  
at proper distance $a^{-1}$ from the horizons, that is the points
\ba \label{conjugate points}
  t_R &=& \frac{1}{a} \sinh a\tau  \, , \qquad t_L = - \frac{1}{a} \sinh a\tau'   \, , 
\nonumber \\
  z_R &=& \frac{1}{a} \cosh a\tau  \, , \qquad z_L = - \frac{1}{a} \cosh a\tau' \, , 
\nonumber \\
  {\bf x}_{\bot \, R} &=& 0 \, , \qquad \qquad\,\,\,\, {\bf x}_{\bot \, L} = 0
\ea
is given by
\ba \label{W_RL}
  {\cal W}_{RL}(\delta)  &\equiv& 
   \bra{0_M} \varphi[x_{R}(\tau)] \, \varphi[x_{L}(\tau')] \ket{0_M} 
  \nonumber \\
   &=& \frac{a^2}{16\pi^2 \cosh^2(a\delta)}
\ea
where $\delta = (\tau - \tau')/2$ (recall that $-\tau'$ is the 
future directed proper time on the UA trajectory in the $L$-wedge), 
see also sec. IV.C in \cite{Obadia:2002qe}.
The correlations are maximal for pairs of conjugate points $\tau = \tau'$ and decay exponentialy
outside the region $\vert \delta \vert \geq 1/a$.
The Fourier transform of the Wightman function w.r.t. $\delta$ is 
\ba \label{TF W_RL}
  \widetilde{\cal W}_{RL}(\lambda) &=& \int_{-\infty}^{+\infty}\!\!d\delta  \, 
     e^{i\lambda \delta} {\cal W}_{RL}(\delta)
\nonumber \\
  &=& - \frac{\vert \lambda \vert}{8\pi} \, \frac{e^{-\pi \lambda/2a}}{1-e^{-\pi \lambda/a}} \, .
\ea
One recognizes the factor $\alpha_\lambda \beta_\lambda$ of the Bogoliubov coefficients
(\ref{Bratio}) and (\ref{alpha}). To show this directly, we use the expansion
of the field in terms of Unruh modes (\ref{phi Unruhmodes}) and evaluate the
expression on the points (\ref{conjugate points}) with the help of the relations
(\ref{phiU})
\ba
  {\cal W}_{RL}(\delta) &=& 2\, \re \int\!\!d^2k_\bot\int_{0}^{+\infty}\!\!d\lambda \, 
  \alpha_\lambda \beta_\lambda \,  \varphi_{\lambda}^{R}(x_R)  \varphi_{\lambda}^L(x_L)
\nonumber \\
  &=&  2 \, \re \int\!\!d^2k_\bot  
 \int_{0}^{+\infty}\!\!\!d\lambda \, e^{-ia\lambda(\tau - \tau')} \, 
  \vert K_{i\lambda}(k_\bot) \vert^2 \, .\qquad 
\ea
The identity
\ba \label{LI K}
  \int_{0}^{\infty}\!\!du \, u \vert K_{i\lambda}(u) \vert^2 
  = \frac{\pi \lambda}{2 \sinh(\pi \lambda)}
\ea
combined with the expressions (\ref{alpha}) and (\ref{Bratio}) finally gives
\ba
 {\cal W}_{RL}(\delta)  = \int_{-\infty}^{+\infty}\!\!\frac{\lambda d\lambda}{4\pi^2}
  \, \frac{e^{-\pi \vert \lambda\vert /a}}{1-e^{-2\pi \vert \lambda\vert /a}} 
  e^{-ia\vert \lambda\vert (\tau - \tau')} \, ,
\ea
whose Fourier transform w.r.t. to $\delta = (\tau - \tau')/2$ is
(\ref{TF W_RL}).

\subsection{$S$-matrix elements}
\label{sec:ampLI}

The $S$-matrix element of the process $\ket{\mp} \to \ket{\pm} + {\bf k}$, i.e.
the exitation (desexitation) of the detector
accompanied by the emission of a particle of momentum ${\bf k}$ 
from the Minkowski vacuum $\ket{0_M}$ are at the lowest order 
\ba\label{Amplitude} 
  {\cal A}_{\pm,{\bf k}}^{LI}
  &\equiv& \bra{0_M} a_{{\bf k}} \otimes \bra{\pm} \: \hat T 
   \: e^{-i\,\int_{-\infty}^{+\infty}\!\! H(\t)} \: \ket{\mp}\otimes\ket{0_M} 
\nonumber \\
  &=& \frac{-i g}{\sqrt{2k(2\pi)^3}}\, 
 \Ie{\t} e^{i[\pm E\t + \omega_k t(\t) - {\bf k}{\bf x}(\tau)]} \, .\quad
\ea
For the inertial trajectories (\ref{Intraj}) we get
\ba \label{AmplInertialLI}
   {\cal A}_{\pm,{\bf k}}^{In, LI} = \frac{-i g}{\sqrt{4\pi k}} \; 
   \delta\left(\omega_k' \pm  E\right) \ ,
\ea
where 
\ba \label{positivity cond}
  \omega_k' = \gamma\left( \omega_k - \beta k_z \right) > 0 
\ea
is the energy of the scalar quantum in the rest frame of the detector.
This quantity is stricktly positive in any Lorentz frame, as a consequence
of the stability of the Lorentz invariant vacuum.
Hence the amplitude to spontaneously emit a quantum
from the ground state vanishes. 
The Dirac distribution $\delta(\omega_k' - E)$ 
in the amplitude ${\cal A}_-$ is the expression of the conservation of energy.
These amplitudes are literaly the expression of the stability of the vacuum.

For the UA trajectory, we choose the scale 
$\kappa = a$ in the coordinate system (\ref{coordR}) 
so that the time coordinate $\eta$ coincides with the 
proper time along the UA trajectory (\ref{UAtraj}), which now coincides with 
the coordinate curve $\zeta=0$. We introduce the shorthand notations
\ba  \label{kpm}
   k_\pm = \omega_k \pm k_z \, 
\ea 
We look for an analytic expression of the integral in
\ba\label{Ampl}
    {\cal A}_{\pm,{\bf k}}^{UA, LI} = \frac{-i g/a}{\sqrt{2\omega_k (2\pi)^3}}  
    \Ie{x} e^{\pm i \frac{E}{a}x + i\left(k_- e^x - k_+ e^{-x} \right)/2} \, . \quad
\ea
We recall that $\omega_k^2 = k_z^2 + {\bf k}_\bot^2 \geq k_z^2$, so that
both $k_\pm$ are positive. We assume first ${\bf k}_\bot \neq 0$.
After the change of variables $y=x+\ln\sqrt{k_-/k_+}$
the amplitude reads
\ba \label{amp1}   
   {\cal A}_{\pm,{\bf k}}^{UA, LI} \propto \left( \frac{k_+}{k_-} \right)^{\pm iE/2a} \, 
   \int_{-\infty}^{+\infty}\!\!dy \, e^{\pm i \frac{E}{a}y - i k_\bot \sinh(y)} \, , 
\ea
We evaluate it as a contour integral along the rectangle 
with edges on $\im(y)=0$ and $\im(y) = -\pi/2$. 
The integrals  along the vertical axis vanish and 
the integral along $\im(y) = -\pi/2$ is the 
integral representation of the modified Bessel function
$2K_\nu(z)=\Ie{y}e^{-z\cosh y+\nu y}, \, \re(z)>0$. We get 
\ba\label{ampUALI}
   {\cal A}_{\pm,{\bf k}}^{UA, LI} 
   = \frac{- i 2g }{\sqrt{2\omega_k (2\pi)^3} }\;
   \frac{e^{\mp \frac{\pi E}{2a}}}{a}    
   \left( \frac{k_+}{k_-} \right)^{\pm i \frac{E}{2a}} 
  K_{\pm i\frac{E}{a}}\left( \frac{k_\bot}{a} \right) \, . \quad
\ea
The case ${\bf k}_\bot = 0$ must be treated separately but the expressions turn out to
be the limiting values of the r.h.s. of (\ref{ampUALI}) \cite{OCa}.
Since $K_{-\nu}(x) = K_{\nu}(x)$, 
the corresponding probabilities 
$P_{\pm,{\bf k}} = \vert {\cal A}_{\pm,{\bf k}}^{UA, LI}  \vert^2$
differ by a Boltzman factor
\ba 
   \frac{P_{+,{\bf k}}}{P_{-,{\bf k}}} = e^{-2\pi E/a} \, .
\ea

The origin of the Boltzmann ratio and the relationship with the 
analytic structure of the Wightman function are 
perhaps better understood from the 
saddle point approximation of (\ref{amp1}). The saddle points
of ${\cal A}_\pm$ are
\ba
   x_n^{+} = {\rm argch}(E/k_\bot) + i n 2\pi \, , \qquad
   x_n^{-} =  i\pi + x_n^{+} \, .
\ea
First, the periodicity of these saddle points reflects the periodicity 
of the poles of the Wightman function. (This should not be 
a surprise because the Wightman function and the amplitudes are 
integrals of the mode functions).
Second, once exponentiated, the relative shift $i\pi$ between the 
positions of the saddle points gives the Boltzmann factor
$e^{-iEx_n^{-}/a} = e^{\pi E/a} e^{-iEx_n^{+}/a}$.
(A more complete analysis can be found in appendix \ref{app:saddle} 
and section \ref{sec:ampLV}).

Last but not least, the $S$-matrix elements are proportional to the Bogoliubov
coefficients between Minkowski and Rindler modes
(these expressions are shown in appendix \ref{app:Bogol}), 
\ba \label{A proto alpha}
  {\cal A}_{-,{\bf k}}^{UA,LI} &=& - i \frac{g}{\pi} 
  K_{i\la/a}\left(\frac{k_\bot}{a}\right)
  \sqrt{\frac{\sinh(\pi\la/a)}{a}} \; \al_{\la,k_z,{\bf k}_\bot}
 \nonumber \\
  {\cal A}_{+,{\bf k}}^{UA,LI} &=&  i \frac{g}{\pi} 
  K_{i\la/a}\left(\frac{k_\bot}{a}\right)
  \sqrt{\frac{\sinh(\pi\la/a)}{a}} \; \beta_{\la,k_z,{\bf k}_\bot}^*\, . \qquad
\ea
This is the consequence of the facts that 
trajectories of UA observers are curves of constant Rindler coordinate $\zeta$, and
along these trajectories $\tau = a \eta$ is proportional to the Rindler time
coordinate. 
In consequence the amplitudes and Bogoliubov coefficients
are essentially given by the same integral.
{\it Lorentz symmetry thus endows the 
Bogoliubov coefficents with a dynamical interpretation.}

\subsection{Transition rates}
\label{sec:ratesLI}

From the $S$-matrix elements of the previous section one forms the inclusive
probabilities
\ba
  P_{\pm} = \int\!\!d^3k \, \left| {\cal A}_{\pm,{\bf k}} \right|^2 
\ea
by summing over the final states of the field.
They can be calculated directly from the 
expressions of the probability amplitudes given in the previous section,
or alternately by exchanging the order of integration over ${\bf k}$ and time.
The calculation is facilitated by keeping the upper bound of the 
time integration in (\ref{Amplitude}) finite and taking the limit $\tau \to \infty$
at the end. Calculations with the first method
can be found in \cite{Matsas}. We adopt the second method \cite{GrandeOeuvre} 
which relates the probability to the Wightman function of the field,
\ba \label{Ppm} 
  P_\pm(\t) = 2 g^2 \re \int^\t_{-\infty}\!\!d\t_1
  \int^{+\infty}_0\!\!d\t_2 \,
  e^{\mp iE\t_2} \: W(\t_1,\t_1-\t_2)  \quad
\ea
where the Wightman function is evaluated at two points on the trajectory of the detector.
The corresponding transition rates of the detector are defined by
\ba \label{Rpm}
  R_\pm(\t) &\equiv& \frac{dP_{\pm}}{d\tau}
\nonumber \\
  &=& 2 g^2 \: \re \int^{+\infty}_0\!\!d\t' \; e^{\mp iE\t'} \: W(\t,\t-\t') \, .
\ea

From the expressions (\ref{WIn}) and (\ref{WUA}), we see that 
$W(\t,\t-\t')$ in the integrand depends only on the 
difference of its arguments $\tau'$ 
both for inertial (\ref{Intraj}) and UA trajectories (\ref{UAtraj}).
The rates $R_{\pm}$ are thus time-independent.
Again, this is because both trajectories are orbits of a Killing vector of Minkowski space
and because the Minkowski vacuum state is annihilated by the corresponding generators.
Namely, inertial trajectories are invariant by time-translation
since $\partial_\tau\vert_{In} = \gamma \partial_t$, and
for UA trajectories a translation along the proper time is a boost since
$\partial_\tau\vert_{ua} \propto t\partial_z + z\partial_t$.

With this simplification, one can write (\ref{Rpm})
as an integral along the entire real line 
\ba \label{FT1}
  R_\pm^{LI} = g^2 \, \re \int_{-\infty}^{+\infty}\!\!d\tau \, e^{\mp i E\tau} {\cal W}(\tau) 
\ea
where ${\cal W}(\tau)$ is either (\ref{WIn}) or (\ref{WUA}).
One can calculate this integral by the method of residues. On inertial trajectories,
the Wightman function (\ref{WIn})
has a double pole a $\tau = i\epsilon$ and 
is analytic in the lower half complex $\tau$-plane (since $\epsilon > 0$), 
which yields
\ba  \label{RIn}
   R_+^{In,LI} = 0 \, , \qquad R_-^{LI,In} = \frac{g^2E}{2\pi}
\ea
The first rate vanishes because of the stability of the vacuum 
in all Lorentz frames.

For UA trajectories, the Wightman function (\ref{WUA}) 
has now a countable family of double poles $\tau_n = i(n2\pi/a + \epsilon)$
periodically spaced on the imaginary axis. As a result
\ba \label{R+acc}
  R_\pm^{UA,LI} &=& \pm \frac{g^2E}{2\pi}\: \inv{e^{\pm \frac{2\pi E}{a}}-1} \ .
\ea
As for the amplitudes, the Boltzmann factor follows directly 
from this periodicity and the stability of the vacuum
(the $i\epsilon$ prescription says that the pole $\tau_0$ counts in $R_-$ but not in $R_+$)
\ba \label{ratiorates}
   \frac{R_+^{UA,LI}}{R_-^{UA,LI}} = e^{-2\pi E / a} \ .
\ea

\subsection{The double role of Lorentz symmetry}
\label{sec:double role}

Lorentz invariance is instrumental in the previous results in two respects.
First, to ensure the stability of the ground state in every frame. The latter is sole 
responsible of the properties of inertial detectors.
This stability implies i) the analyticity and boundedness of the 
Minkowski modes (\ref{Minkmodes}) on an inertial trajectory in the domain 
$\im(\tau) < 0$.
This in turns implies ii) the analyticity of the Wightman function in $\im(\tau) < 0$,
iii) ${\cal A}_+^{In} = 0$ and iv) $R_+^{In} = 0$. 

\begin{widetext}
These four properties are in fact equivalent. Three of the equivalences can readily 
be shown directly: 
i) $\Leftrightarrow$ ii) because the measure $d^3k/\omega_k$
is Lorentz invariant; i) $\Leftrightarrow$ iii) and 
ii) $\Leftrightarrow$ iv) by Fourier transform (\ref{Ampl}) and (\ref{FT1}) respectively.
One can also show directly that the transition amplitudes give back the transition rates, 
again thanks to the Lorentz invariant measure.
\ba
\begin{array}{ccc}
\mbox{ i)} \, \varphi^M_{\bf k}(x^a_{In}(\t)) \,\,  \mbox{analytic \& bounded }& 
  \stackrel{LI \,\, of \,\, d^3k/2\omega_k}\Longleftrightarrow & 
  \mbox{ ii)} \,  \mbox{Analyticity of } {\cal W}_{In}(\tau) \\
& & \\
 FT \Updownarrow &   &  \Updownarrow FT \\
& & \\
  \mbox{ iii)} \,\left\{
\begin{array}{l}
{\cal A}_{+,{\bf k}}^{In,LI} = 0 \\
{\cal A}_{-,{\bf k}}^{In,LI} \neq 0
\end{array}
\right.
& \stackrel{LI \,\, of \,\, d^3k/2\omega_k}{\Longrightarrow} &
  \mbox{ iv)} \,\left\{
\begin{array}{l}
R_+^{In,LI} = 0 \\
R_-^{In,LI} \neq 0
\end{array}
\right.
\\
\end{array}
\nonumber
\ea
\end{widetext}

The second role of Lorentz symmetry pertains to the UA trajectories (\ref{UAtraj}), which
are orbits of the generator of boosts along the direction $z$, and are periodic in 
$\im(\tau)$. The first property implies stationarity, and the second property 
combined with the stability of the vacuum gives the thermal spectrum.
a) These properties correspond respectively 
to eqs. (\ref{phiU}) and (\ref{CL mink}) of the definition of the Unruh modes,
the requirement of analyticity fixing the ratio of the Bogoliubov coefficients, see
eqs (\ref{asymp phiU}) and (\ref{Bratio}). 
Stationarity and stability are also transparent on 
b) the expression (\ref{WUA}) of the Wightman function ${\cal W}_{ua}$.
These properties are directly inherited from the Unruh (or Rindler) modes
by  Lorentz invariance of the measure $d^3k/\vert {\bf k}\vert$, this time 
expressed in the form of eq. (\ref{LI K}).
By Fourier transform, 
the periodicity of poles of the Wightman function implies that c) both $R_\pm$ are
proportional to the Planckian spectrum $\left(\exp(2\pi E/a) - 1\right)^{-1}$, 
and the stability of the Minkowski vacuum fixes the ratio of the rates (the pole $+i\epsilon$
contributes to $R_-$), see eqs. (\ref{R+acc}) and (\ref{ratiorates}).
Similarly, as seen on (\ref{amp1}), the periodicity of the UA trajectory implies 
that d) $\vert {\cal A}_+ \vert \propto \vert {\cal A}_- \vert$, and the stability of the 
vacuum fixes the ratio (this will be clearer in sec. \ref{sec:ampLV} by comparison 
with the amplitudes for subluminal dispersion relations). 
Finally, the fact that UA trajectories are orbits of the boost generator 
is responsible for e) the proportionality of 
the Bogoliubov coefficients with the transition amplitudes 
eq. (\ref{A proto alpha}), which gives the 
former a physical interpretation.  
\begin{widetext}
Properties a)-d) are also equivalent by Lorentz invariance: 
from eq. (\ref{Bratio}), (\ref{egality Bogol}) and (\ref{A proto alpha}) we have 
a) $\Leftrightarrow$ d); 
a) $\Leftrightarrow$ b) because of the Lorentz invariant measure $d^3k/\vert {\bf k}\vert$;
b) $\Leftrightarrow$ c) by Fourier transform;
and by Lorentz invariance of the measure, one shows \cite{Matsas} that d) $\Rightarrow$ c).
\ba
\begin{array}{cccc}
  \mbox{ a)}\, \varphi^U_{\Omega} \,\,  \mbox{analytic \& bounded } 
  & \stackrel{LI \,\, of \,\, d^3k/2\omega_k}\Longleftrightarrow 
  & \mbox{ b)} \,\mbox{Poles of} \,\, {\cal W}_{ua}(\tau) \\
  & & & \\
 eq. (\ref{A proto alpha}) \Updownarrow &   &  \Updownarrow FT \\
& & & \\
 \mbox{ d)} \,\left|{\cal A}_{+,{\bf k}}^{UA,LI}/{\cal A}_{-,{\bf k}}^{UA,LI}\right|^2 
      = e^{-2\pi E/a}
& \stackrel{LI \,\, of \,\, d^3k/2\omega_k}{\Longrightarrow} &
\mbox{ c)} \,R_+^{UA,LI}/R_-^{UA,LI} = e^{-2\pi E/a}\\
\end{array}
\nonumber
\ea
\end{widetext}
In a nutshell, both roles of Lorentz symmetry are {\it sufficient} 
to the existence of the Unruh effect. 
The effect can be characterised by either of the four properties a)-d) presented 
in sections \ref{sec:stateLI}-\ref{sec:ratesLI} because they are 
equivalent by Lorentz invariance.
We will now show that Lorentz symmetry is also {\it necessary} 
for the existence of the Unruh effect.

\section{No Unruh effect without Lorentz invariance}
\label{sec:wihtoutLI}

Whenever possible, we will establish general results valid for 
arbitrary dispersion relations (DR) which we note $\omega_k$. 
The phase and group velocities will be noted
\ba
   v_\varphi \equiv \frac{\omega_k}{k} \, , \qquad 
   v_g \equiv \frac{d \omega_k}{dk} \, .
\ea
We recall that we work in the units where the
velocity of light is $1$, which is the asymptotic velocity of the UA trajectory
(\ref{UAtraj}). The DR is called subluminal (possibly on
a finite interval only) if $v_\varphi < 1$, and superluminal if $v_\varphi > 1$.
We will illustrate our results with the particular case of linear dispersion relations
\ba  \label{linDR}
   \omega_k = v k \, , \qquad v \neq 1 \, .
\ea
This case can be treated to a large extend analytically because
the field still enjoys a Lorentz symmetry. 
The state of the field is the ground state defined in the preferred frame
and noted ${\ket{0}}$ (the pseudo-Minkowski vacuum).

\subsection{Bogoliubov coefficients}
\label{sec:Bogol IV}

We first extend the construction of \ref{sec:stateLI} to linear DR (\ref{linDR})
and show explicitly that the density matrix $\rho$ defined at equation (\ref{rho})
is not thermal. The sub- and superluminal cases must be treated separately.
The construction of the Unruh modes is 
identical to the relativistic case for superluminal DR but it must be amended for $v<1$
because of the instability of the vacuum w.r.t. frames $\beta > v$.
We then briefly discuss generalizations to arbitrary DR.

\subsubsection{Linear dispersion relations}

The starting point of the analysis of sec. \ref{sec:stateLI} is 
the observation that the $R$-wedge is globally hyperbolic, 
or equivalently a smooth time coordinate
(namely $\eta$ from eq. (\ref{coordR})) can be chosen in $R$ such that
the surfaces $\eta = {\rm cte}$ are Cauchy surfaces.
This is a necessary condition so that the solutions of the wave equation
$\Box \varphi = 0$ are uniquely determined by boundary conditions 
on that Cauchy surface.
This can be easely adapted to the linear DR (\ref{linDR}).
The wave equation in the preferred frame is
\ba \label{eom DR}
   \left(\di_t^2-v^2(\di_x^2 + \di_y^2 + \di_z^2)\right) \varphi &=& 0
\ea
which is trivially put into the form (\ref{eom Mink}) by
a rescalling $t' = vt$
since the mass-shell relation (\ref{linDR}) enjoys an $SO(1,3)$ symmetry. 
The causal properties (\ref{eom DR}) are determined by the pseudo-light cones
$v \vert t \vert = r$, so we define pseudo Rindler wedges
$\tilde R \, (\tilde L) = \left\{ (t,x,y,z) / z \gtrless 0 , v \vert t \vert < \pm z \right\}$. 
The metric in these globally hyberbolic space-times is brought into
static form by the introduction of the pseudo-Rindler coordinates
\ba
  t = \frac{1}{v} \, e^{\tilde\zeta} \sinh(\tilde\eta)   
  \,  , \qquad 
  z =  \, e^{\tilde\zeta} \cosh(\tilde\eta)
\ea
in $\tilde R$ and with a minus sign in $\tilde L$.
We can proceed with the quantization following sec. \ref{sec:stateLI} step by step
up to eq. (\ref{PhiRindler}) and the definition of the 
a pseudo-Rindler vacuum $\ket{\tilde 0_R} \ket{\tilde 0_L}$.

\begin{figure}[h!]
\hbox to\linewidth{\hss
\resizebox{7cm}{7cm}{\includegraphics{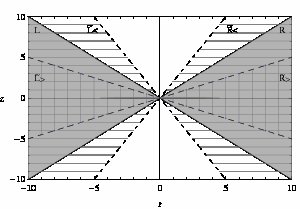}}\hss}
\caption{The Rindler wedges $\tilde R$ and $\tilde L$ for subluminal (subscript $<$)
and superluminal (subscript $>$) linear dispersion relations (\ref{linDR}).
\label{diagrMink}}
\end{figure}

At this point we must distinguish between super and subluminal dispersion relations.
If $v>1$, the energy of the pseudo-Minkowski modes is in the preferred frame
\ba \label{positivity}
 \omega_k^2 = v^2 \left( k_z^2 + {\bf k}_\bot^2 \right) \geq v^2 k_z^2 > k_z^2 \, .
\ea
The pseudo-Minkowski modes are therefore analytic and bounded functions
over the same domain ${\cal T}$ as the Minkowski modes. The pseudo-Minkowski 
vacuum is stable, i.e. this is the lowest energy state in any inertial frame.
We can therefore continue to follow the procedure of sec. \ref{sec:stateLI}
step-by-step, define Unruh modes, which gives the unitary map between the 
pseudo-Minkowski vacuum and pseudo-Rindler Fock state (\ref{map}).
The point of departure with sec. \ref{sec:stateLI} 
is that we assume that the observers live 
in the wedge $R = \left\{ z > 0, \vert t \vert < z \right\}$.
If the DR is superluminal, $\tilde R \subset R$ and 
there exists observables such that ${\rm supp}({\cal O}_R) \subset R \backslash \tilde R$
where the pseudo-Rindler vacuum $\ket{0}_{\tilde R}$ is not defined, 
see fig. (\ref{diagrMink}).
The trace over $L$ of $\ket{\tilde 0_M}\bra{\tilde 0_M}$ is therefore not given by
(\ref{rhoth}).

If the DR is subluminal, modes with transverse wavenumbers small enough to verify
\ba
  k_\bot^2 < \frac{1-v^2}{v^2} k_z^2
\ea
break the positivity condition (\ref{positivity}). For those modes, we must require
that $\varphi^U_\Omega$ of eqs. (\ref{phiU}), 
restricted on the future pseudo-horizon $\widetilde{\cal H}^+$, be analytic and bounded 
in $\im(\widetilde V) > 0$, where $\widetilde V = vt + z$.
The unique analytic continuation is $(\widetilde V + i\epsilon)^{i\Omega}$, which
leads to an inversion of the ratio 
$\tilde \beta_\Omega / \tilde \alpha_\Omega = e^{+\pi \Omega}$.
(This means not only that for observables restricted to this $\tilde R$ wedge, 
the density matrix is not thermal, but that it is not a trace operator since 
${\rm Tr}(\rho) = \infty$.)
Then, as in the superluminal case we assume that the observers live 
in the wedge $R = \left\{ z > 0, \vert t \vert < z \right\} \subset \tilde R$
and similarly $L \subset \tilde L$.
The state resulting from tracing over $L$ cannot be thermal because it is 
still correlated in the region $\tilde L \backslash L$.

In conclusion, the pseudo-Minkowski vacuum restricted to the Rindler wedge 
$R$ is not a thermal state.
The Bogoliubov coefficients $\tilde \alpha$ and $\tilde \beta$ 
give the expression of the pseudo-Minkowski vacuum in terms of the 
pseudo-Rindler quanta, whereas we are interested in the expression of the 
pseudo-Minkowski vacuum in terms of the Rindler quanta, i.e.
\ba
  \bra{\tilde 0_M} a_{\lambda}^{R\,\dagger}  a_{\lambda'}^{R\,\dagger} \ket{\tilde 0_M} 
  \neq 
 \delta(\lambda - \lambda')\, \vert \tilde \beta_{\lambda} \vert^2
\, , \nonumber \\
  \bra{\tilde 0_M} a_{\lambda}^{R}  a_{\lambda'}^{L} \ket{\tilde 0_M} 
  \neq  \delta(\lambda - \lambda')\,\tilde \alpha_{\lambda}\tilde  \beta_{\lambda}   \, .
\ea
We will see in sec. \ref{sec:ampLV} that the second interpretation of the 
Bogoliubov coefficients in terms of $S$-matrix elements is
also lost.

\subsubsection{Generalizations}

In the more general case $\omega^2 = F(k^2)$, the mode equation in the 
preferred frame is given by
$\left[ \partial_t^2 - F(\partial^2) \right] \varphi(x) = 0$.
Since it lacks an $SO(1,3)$ symmetry, the wave equation in Rindler coordinates
mixes non-linearly time and space derivatives, and the notion of  
Rindler wedges $\tilde R,\, \tilde L$ looses its significance. 

But since the settings for the Unruh effect pick a preferred direction, 
see eqn. (\ref{UAtraj}), one could be curious to consider 
instead anisotropic dispersion relations with an $SO(1,1)$ symmetry
\ba \label{SO(1,1)}
  \omega^2(k_z,{\bf k}_\bot) = k_z^2 + F({\bf k}_\bot^2) \, . 
\ea
This is of course very contrived physically, but we will see that 
formally this mimics
rather closely the settings of Hawking radiation.
The pseudo-Minkowski modes solutions of (\ref{SO(1,1)}) 
are analytic and bounded in ${\cal T}$ if and only if
$F\geq 0$.
We can go on defining Rindler coordinates in which the wave equation 
takes the form
\ba \label{contrived}
  \left( \partial_\eta^2 - \partial_\zeta^2 + e^{2\zeta} F(\partial_\bot^2) \right)  
  \varphi = 0 \, .
\ea
The solutions of positive Rindler frequency are given by (\ref{phiR})
with $k_\bot e^\zeta$ replaced by $\sqrt{F(k_\bot^2)} e^\zeta$ 
in the argument of the Bessel function. We see again that provided $F \geq 0$,
the properties of the mode functions are not altered w.r.t. the Lorentz invariant case
and we can proceed in a similar fashion as in sec. (\ref{sec:stateLI}). 
In that case we thus recover the Unruh effect at the level of the Bogoliubov coefficients,
i.e. eqs. (\ref{asymp phiU})-(\ref{alpha}) hold. 
This is because the characteristics $t = \pm z$ of eq. (\ref{contrived}) 
generate the Rindler horizons ${\cal H}^{\pm}$. We shall return to this in 
sec. \ref{sec:compare}.

\subsection{Wightman function}
\label{sec:W LV}

We examine how modifying the dispersion relation affects the analytical properties 
of the Wightman function. In brief: 1) the analytic properties are essentially the 
expression of the stability or instability of the vacuum in all inertial frames. 
There is therefore a sharp distinction between super- and subluminal DR, the latter
defining pathological models. 2) Both properties of the Wightman function 
evaluated on UA trajectory
(\ref{WUA}), that is invariance by boost 
and the equilibrium condition (\ref{KMS}) are lost.

\subsubsection{Inertial frames}

In the preferred frame, the solutions of the wave equation
$\varphi_k \propto e^{-i\omega_kt + i{\bf k}{\bf x}}$ have a positive Klein-Gordon norm.
In a boosted frame $t'=\gamma(t+\beta z), \, z'=\gamma (z + \beta t)$, 
the norm of the modes 
$\varphi_k \propto \exp\left( -i \omega_k't' + i k'z' \right)$
now depends on the sign of the boosted frequency $\omega_k' = \gamma (\omega_k - \beta k_z)$
\ba
  \left( \varphi_{k'} ,\, \varphi_k \right) \propto {\rm sgn}(\omega_k') 
  \, \delta^{(3)}({\bf k} - {\bf k}') \, .
\ea
We must thus distinguish two cases. 
First, with superluminal dispersion relations
\ba \label{condpositivity}
  \forall k \, , \quad  v_{\varphi} \geq 1  \, .
\ea
solutions of positive frequency (or norm) in the preferred frame have
positive frequency in all frames. The vacuum $\ket{0}$ defined in the preferred frame is 
therefore the ground state in all the frames.
Hence the Fourier transform of the Wightman function 
$W = \bra{0} \varphi(x+y) \varphi(y) \ket{0} = \int \varphi_{k}^* \varphi_{k}$ 
contains only positive frequencies in any frame, and provided the phase velocity does 
not vanish (otherwise the factor $1/2\omega_k$ introduces poles or branch cuts), 
$W(x)$ has the same analytical properties as in the relativistic case
in the sense that the regulator $e^{-\epsilon k}$ is equivalent to the replacement
$t \mapsto t-i\epsilon$.
If the DR violates the positivity condition (\ref{condpositivity}), 
the Fourier transform of $W$ contains negative frequencies 
w.r.t. the modes defined in boosted frames such that $1 > \beta > v_\varphi$
and the analytic properties of $W$ are changed.

Let us illustrate this with the linear dispersion relations (\ref{linDR}).
The unregularized Wightman function is
\ba \label{Wdk}
   {\cal W}_{In}(\tau)
   = \frac{-i}{8\pi^2\gamma\beta v \vert \t \vert} \int_0^{\infty}\!\!dk \;
   \left( e^{-i\gamma k(v -\beta \eta_\tau) \tau }   \right.
\nonumber \\ -
   \left.  
   e^{-i \gamma k(v + \beta \eta_\tau) \tau}\right) \, ,
\ea
where we note $\eta_\tau = {\rm sgn}(\tau)$.
The absolute values come from $r = \vert z \vert$, see eq. (\ref{Wint}).
As we said, if $v> \vert \beta \vert$ we can regularize the integrand 
with $e^{-\epsilon k}$, or equivalently replace $\tau$ by $\tau-i\epsilon$.
The integration is straightforward,
\ba \label{Wok}
  {\cal W}_{In}(\tau) &=&  -\frac{1}{4\pi^2 v} \frac{1-\beta^2}{v^2 - \beta^2} \, 
  \frac{1}{\tau(\tau-i\epsilon)} 
\ea
and one can write the $\tau$-dependent part in three equivalent ways
\ba
   \frac{1}{\tau(\tau-i\epsilon)} &=& \frac{1}{\tau^2 - i\eta_\tau \,\epsilon} 
   = \frac{1}{(\tau - i\epsilon)^2} 
\ea
The third expression shows explicitely that 
${\cal W}_{In}$ is analytic in the lower half complex $\tau$-plane.
For subluminal DR and frames such that $\beta < v < 1$, this remains true.
If $v< \beta <1$ on the other hand, the Wightman function
admits a different representation
\ba
  {\cal W}_{In}(\tau) = 
   \frac{-i }{8\pi^2v\gamma\beta \vert \t \vert} \int_0^{\infty}\!\!dk \;
   \left( e^{-i\gamma k (v-\beta\eta_\tau) (\t +i\eta_\tau\epsilon)}  \right.
\nonumber \\ -
   \left.  
   e^{-i \gamma k (v+\beta \eta_\tau)  (\t-i\eta_\tau\epsilon)}\right)
\ea
the integration of which gives
\ba \label{Wcata}
  {\cal W}_{In}(\tau) 
   &=& -\frac{1}{4\pi^2v} \frac{1-\beta^2}{v^2 - \beta^2} \frac{1}{(\tau + i\epsilon)^2}
\, .
\ea
It is the complex conjugate of the (\ref{Wok}) and 
is analytic in the upper half complex $\tau$-plane.
As a result, $R_-$ will be found to vanish instead of $R_+$.

\begin{widetext}
\subsubsection{UA frames}
The same considerations apply to the Wightman function on a UA trajectory.
Substituting the parametrization (\ref{UAtraj}) into the 
regularized integral expression of the Wightman function gives
\ba\label{bloodyW}
  {\cal W}_{ua}(\delta,\bar \tau) &=& \frac{-i}{8\pi^2 \vert \Delta z \vert} 
  \int_0^{\infty}\!\! \frac{dk}{v_\varphi} \,
  \left\{   e^{-i2\frac{k}{a}\sinh(a\delta/2)\left( 
   v_\varphi \cosh(a\bar \tau) - \eta_\delta \vert \sinh(a\bar \tau) \vert \right) } 
  - e^{-i2\frac{k}{a}\sinh(a\delta/2)\left( 
   v_\varphi \cosh(a\bar \tau) + \eta_\delta \vert \sinh(a\bar \tau) \vert \right) } 
 \right\}e^{-\epsilon k}\qquad
\ea
\end{widetext}
We introduced the notations 
\ba
  \delta = \tau_1 - \tau_2 \, ,\qquad \bar \tau = \frac{\tau_1 + \tau_2}{2}
\ea
and $\Delta z = z(\tau_1) - z(\tau_2) = 2 \sinh(a\delta/2) \sinh(a\bar \tau)$.
For superluminal DR, that is satisfying (\ref{condpositivity}), 
the prefactor $v_\varphi \cosh(a\bar \tau) \pm \eta_\delta \vert \sinh(a\bar \tau)\vert $ 
in the first phase
is stricktly positive, so that $e^{-\epsilon k}$ can be replaced by
$\delta \to \delta - i\epsilon$ since
\ba \label{absorb epsilon}
  \sinh(x-i\epsilon) &=& \sinh(x) \cos(\epsilon) -i\cosh(x) \sin(\epsilon)
\nonumber \\
   &=& \sinh(x) - i \epsilon \, \cosh(x) + O(\epsilon^2)
\nonumber \\
   &\to&  \sinh(x) - i\epsilon 
\ea
If the DR is subluminal at some value $k$, 
the function $v_\varphi(k) \cosh(a\bar \tau) \pm \eta_\delta  \vert \sinh(a\bar \tau) \vert$
changes sign at $\tau(k)$ given by 
\ba 
  \tanh\left( a \vert \tau(k) \vert \right) \equiv v_\varphi(k)
\ea
We cannot replace $e^{-\epsilon k}$ by a single prescription $\delta \pm i\epsilon$.

The linear DR (\ref{linDR}) provides again a good illustration of this.
Once the integration in (\ref{bloodyW}) done and the fractions combined, we have
\ba 
  &&{\cal W}_{ua}(\delta,\bar \tau) = -\frac{a^2}{16\pi^2} \frac{1}{\sinh^2(a\delta/2)} \times
 \nonumber \\
 &&\frac{1}{\left[ v^2\cosh^2(a\bar \tau) - \sinh^2(a\bar \tau)\right]
   - i \epsilon\, 2a \Delta t} \qquad
\ea
with $a \Delta t = 2\sinh(a\delta/2) \cosh(a\bar \tau)$.
If $v > 1$, we can replace the second denominator by
\ba
   \left[ v^2\cosh^2(a\bar \tau) - \sinh^2(a\bar \tau) \right]\times \left( 1 - i\epsilon \right)
\ea
and finally absorb the $i\epsilon$ into $\sinh^2(a\delta/2)$ as done in 
(\ref{absorb epsilon}). Hence
\ba \label{WLV}
  v > 1 \, , \quad 
  {\cal W}_{ua}(\tau_1,\tau_2) = {\cal W}_{\rm LI}(\delta) \times 
  f\left(\bar \tau \right)
\ea
where the Lorentz invariant Wightman function 
${\cal W}_{\rm LI}(\delta)$ is given at eq. (\ref{WUA})
and the function $f$ is given by
\ba \label{def f}
  f(\bar \tau) = \frac{1}{v} \, \frac{1}{v^2 \cosh^2(a\bar \tau) - \sinh^2(a\bar \tau)}
\ea
Because of $f$, ${\cal W}_{ua}$ is not stationary, and a fortiori not thermal.
We stress that the factorization of the dependences in $\delta$ and $\bar \tau$ 
is not generic. 
It is a consequence of the linearity 
of the dispersion relation, and of the properties of the UA trajectories.

For subluminal DR, the function $v^2 \cosh^2(a\bar \tau) - \sinh^2(a\bar \tau)$
becomes negative at the time $\rho$ 
given by $\tanh(a\rho) = v$. Thus for $\bar \tau \leq \rho$ we can still use
the prescription $\delta -i\epsilon$, but for 
$\bar \tau \geq \rho$ we must replace it by $\delta + i\epsilon$.

\subsection{Transition rates}
\label{sec:RLV}

In the Golden rule limit (\ref{Rpm}), the transition rates are given by a Fourier 
transform of the Wightman function. The properties of the latter therefore
pass on to the former:
the transition rates are not stationary, and their ratio is not
the Boltzmann factor. We show this explicitly with linear DR (\ref{linDR}).
In appendix \ref{app:Taylor} we use asymptotic expansions of the Wightman function
and transition rates to estimate the proper time interval elapsed since the beginning
of the acceleration after which stationarity and thermality are lost for more
general DR. The result is
\ba
   2\pi \ll a\tau \ll 
  \frac{1}{2}\ln\left\{ {\rm min}\left(\frac{M}{E}, \frac{M}{a} \right) \right\}
\ea
The lower bound excludes transients.
If we take $M$ equal to the Planck mass, 
the upper bound can be as high as $100$.

\subsubsection{Inertial detector}
\label{sec:RLV inertial}

We start with the integral representation (\ref{Wdk}) regularized with $e^{-\epsilon k}$
and integrate by parts
\begin{widetext}
\begin{eqnarray}
   {\cal W}_{In}(\tau)
   = \frac{-i}{8\pi^2\gamma\beta \t} 
    \left[G(k)\left( e^{-i \omega_- \t} 
     -e^{-i \omega_+ \t}\right)e^{-\epsilon k}\right]_0^\infty 
   + \inv{8\pi^2\beta} \int_0^{\infty}\!\!dk \, G(k)
    \left\{ (v_g-\beta) e^{-i \omega_- \t}
    - (v_g + \beta) e^{-i \omega_+ \t}\right\} \,e^{-\epsilon k} \quad
\end{eqnarray}
where $G(k)=\int^k\!\!\frac{dk}{v_\varphi}$ 
and we note $\omega_{\pm} = \gamma(\omega_k \pm \beta k)$.
The boundary term vanishes at both the lower and upper bounds.
We then substitute this expression in (\ref{Rpm})
and exchange the order of integrations 
\ba
   R_\pm^{In}  &=& g^2 \: \int^{+\infty}_{-\infty}\!\!d\t \; e^{\mp iE\t} \: {\cal W}_{In}(\t) 
\nonumber \\
    &=& \frac{g^2}{4\pi\beta} \:  \int_0^{\infty}\!\!dk \, G(k) \left\{ (v_g-\beta) 
   \delta(\omega_- \pm E) 
 - (v_g+\beta ) \delta(\omega_+ \pm E) \right\} \,e^{-\epsilon k} \, ,
\ea
\end{widetext}
As expected, $R_+^{In}  \neq 0$ if $\omega_- < 0$.
We can further write
\ba\label{RinCG}
   R_{-}^{In} =  \frac{g^2}{4\pi\beta\gamma} \: 
   \sum_p \left( s_p^- G(k_p^-) - s_p^+ G(k_p^+) \right)
\ea
where $k^{\pm}_{p}$ and $s_p^\pm$ are defined by 
\ba
  E &=& \omega_\pm(k^{\pm}_{p}) \, , \qquad
  s_p^\pm = \mbox{sgn}(v_g(k_p^\pm)\pm\beta)
\ea

The calculation with a linear dispersion relation (\ref{linDR})
provides an independent check of this result since in that
case we have integrated over $k$ first, and the integral over $\tau$
can be done by application of the theorem of residues.
For superluminal DR, we use (\ref{Wok}) and get
\ba
  R_+ = 0  \, , \qquad 
  R_- &=& \frac{g^2 E}{2\pi} \frac{1-\beta^2}{v (v^2 - \beta^2)}
  \, , 
\ea
which is also the result obtained by application of the expression (\ref{RinCG}).

For subluminal DR, the previous result is still true as long a $\beta < v$.
For $\beta = v$ the Wightman function is not defined and for $\beta > v$
the $i\epsilon$ prescription must be complex conjugated, i.e.
the numerator of (\ref{WIn}) is $(v^2 -\beta^2)(\tau + i\epsilon)^2$ 
and we now find
\ba
  R_-=0 \, , \qquad R_+ =  \frac{g^2 E}{2\pi^2} \frac{1-\beta^2}{v (\beta^2 - v^2)}
 \, , 
\ea
which matches with (\ref{RinCG}).  
The vacuum of the quantization in the preferred frame 
appears to this observer as a negative energy state
with respect to his ground state.

\subsubsection{UA trajectories}

We lack analytical tools to study the general case
of an arbitrary dispersion relation. 
In appendix \ref{app:Taylor} we use Taylor expansions to show that
stationarity is lost after a few thermal periods $a/2\pi$ at best. 
Below we establish results for the linear dispersion relations (\ref{linDR}). 
The Wightman function (\ref{bloodyW}) is not stationary, 
which means that we cannot
calculate the transition rates (\ref{Rpm}) with a contour integral.
We find however that the "even" part of the transition rates
\ba \label{barR}
  \bar R_{\pm}(\tau) &\equiv& \frac{1}{2} \left\{ R_{\pm}(\tau) + R_{\pm}(-\tau) \right\} 
\nonumber \\
   &=& g^2 Re \int_{-\infty}^{\infty}\!\!d\tau_1 \,  e^{\mp iE\tau_1} W(\tau,\tau-\tau_1)
\ea
can still be calculated as a sum of residues.
We call it the "mean rate". If $\bar R$ is not constant, then $dR/d\tau$ is not an
even function and $R(\tau)$ is not constant.

\begin{figure}[h!] 
\includegraphics[width=7.6cm,height=4.7cm]{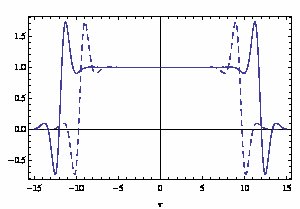}
\caption{Dependence of $\bar R_+(\tau)/R_+^{LI}$ on the velocity. The two curves correspond to
$v=1+10^{-10}$  (plain) and $v=1+10^{-8}$ (dashed) 
with $a=1$, and $E=1$. The deviations from the Lorentz invariant occurs at times 
$\vert \tau \vert \geq \rho$ where $\rho$ is defined at eq. (\ref{defrho}).
Its values are respectively $\simeq 11.9$ and $\simeq 9.6$
\label{fig2}}
\end{figure}

\begin{figure}[h!]
\hbox to\linewidth{\hss
\resizebox{7.6cm}{4.7cm}{\includegraphics{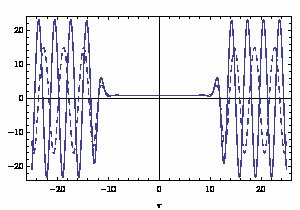}}
\hss}
\caption{The ratio $(\bar R_+(\t)/\bar R_-(\t)) e^{2\pi E/a}$ of the mean rates
normalized to the Boltmann factor. 
We took $v=1+10^{-10}, a=1$, and $E=1$ (plain), $E=0.8$ (dashed).
Deviations from the Lorentz invariant result are of order $1$.
\label{fig3}}
\end{figure}

We must as before distinguish between super and subliminal DR.
We consider first phase velocities $v > 1$. 
We introduce the Rindler time $a\rho$ defined by
\ba \label{defrho}
  \tanh(a\rho) = v  
\ea
in order to write the denominator in (\ref{def f}) as a product
$v^2 \cosh^2(a\bar \tau) - \sinh^2(a\bar \tau)
  = (v^2-1) \, \cosh a(\rho -\bar  \tau) \, \cosh a(\rho + \bar \tau)$.
After the change of variable $x=\frac{a\tau_1}{2}$, the expression of 
$\bar R_+$ is
\begin{widetext}
\begin{eqnarray}
 \bar R_+ &=& -\frac{g^2}{4\pi^2} \frac{a\sinh^2(a\rho)}{2v} \, 
  \re \int_{-\infty}^{+\infty}dx \,
  \frac{e^{i\frac{2E}{a}x}}{\sinh^2\left( x-i\epsilon  \right)} 
 \, \frac{1}{\cosh\left[x- a(\tau-\rho)  \right] \, 
  \cosh\left[x- a(\tau+\rho)  \right] }
\end{eqnarray}
${\cal W}_{\rm LI}$ has a familly of douple poles at $in\pi + i\epsilon$ and $f$ has 
two families of simple poles
\ba \label{poles v>1}
  x_p &=& a(\tau-\rho) + i\frac{\pi}{2} + ip\pi \, \, , \qquad
  y_p = a(\tau+\rho) + i\frac{\pi}{2} + ip\pi
\ea
Note that these poles are shifted by $i\frac{\pi}{2}$ with respect to 
the double poles $in\pi$. The importance of this structure will 
be better understood in the calculation of the amplitudes.
One obtains
\begin{eqnarray}\label{bar R ua}
  \bar R_\pm(\tau)  
   &=& R_\pm^{LI} f(\tau)
   + \frac{g^2a}{16\pi^2 v^2\sinh\left(\pi E/a\right)}
   \left(\frac{\sin[2E(\tau + \rho)]}{\cosh^2a(\tau+\rho)}- 
  \frac{\sin[2E(\tau-\rho)]}{\cosh^2 a(\tau + \rho)}\right)
\end{eqnarray}
\end{widetext}
where $f(\tau)$ is defined at eq. (\ref{def f}).
We refer to the figures \ref{fig2} and \ref{fig3} 
for further details on these expressions.

\begin{figure}[h!]
\hbox to\linewidth{\hss
\resizebox{7.6cm}{4.7cm}{\includegraphics{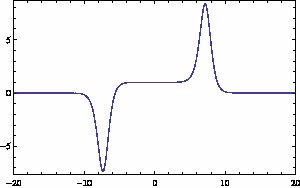}}
\hss}
\caption{The transition rate $R_+(\t)$ 
normalized to the Lorentz invariant value. 
We took $v=1+10^{-6}, a=1,\al=0$, and $E=0.1$. 
It is equal to $1$ for times $\vert \tau \vert \leq \rho \simeq 7.3$, and 
after a burst around $\vert \tau \vert \simeq \rho$ it vanishes exponentially.
\label{plotR}}
\end{figure}

\begin{figure}[h!]
\hbox to\linewidth{\hss
\resizebox{7.6cm}{4.7cm}{\includegraphics{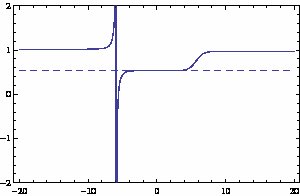}}
\hss}
\caption{The ratio $R_+(\t)/R_-(\t)$. The dashed line is the Boltmann factor. 
We took $v=1+10^{-6}, a=1$, and $E=0.1$.
Interestingly the rates are equal for times $\vert\tau \vert > \rho$.
\label{plotRratio}}
\end{figure}

Although $\bar R_\pm(\t)$ is not the transition rate,
the fact that the second term (\ref{bar R ua}) is common to both mean rates  
shows that the rates are not time independent and that their ratio 
is not a Boltzmann factor for times $\vert \tau \vert \geq \rho$
\ba
  \frac{R_+}{R_-} \neq \exp\left( - \frac{2\pi E}{a} \right) \, .
\ea
We also calculated numerically the transition rates. They are shown on
fig. \ref{plotR} and \ref{plotRratio}. We observe that they are equal to the 
Lorentz invariant ones inside the interval $\vert \tau \vert \leq \rho$, and 
asymptote rapidly to zero outside, but are equal. Between 
these two regimes, the transition rates experience a burst.


For $v< 1$, the Wightman function does not have a single analytic expression 
in $\delta$ for all values of $\tau$ and we cannot calculate $\bar R_{\pm}$
by an integral contour. 
In preparation of the analysis of the $S$-matrix elements, 
it is nevertheless useful to 
examine the changes in the analytic structure of the integrand 
(\ref{barR}) compared to the superluminal case.
For $v<1$, we introduce in place of (\ref{defrho})
\ba
  \coth(a\rho) = v \, .
\ea
The rapidity $a\rho$ was previously introduced after eq. (\ref{def f}) as
the critical value of proper time after which the prescription in the Wightman
function must be changed to $\delta + i\epsilon$.
The denominator of $f$ is now 
$(1- v^2) \sinh a(\rho - \bar \tau) \, \sinh a(\rho + \bar \tau)$,
whose poles are
\ba \label{poles v<1}
    x_p' &=& a(\tau-\rho)  + ip\pi
\nonumber \\
    y_p' &=& a(\tau+\rho)  + ip\pi
\ea
Contrary to superluminal velocities, all the poles are arranged on lines
$\im(\tau) = ip \pi$.
We will now see how this analytic structure appears in the transition amplitudes.

\subsection{$S$-matrix elements}
\label{sec:ampLV}

We first give exact expressions. 
Again sub- and superluminal DR give different results. 
The ratio of the transition probabilities is the Boltzmann factor if
$\omega_k \geq k$, and one if $\omega_k < k$. 
The latter result is a dramatic consequence of 
the instability of the vacuum.

We then calculate again the transition amplitudes by the method of 
steepest descent for the role of the 
analytic properties of the modes in the previous results
appears more clearly with this method.

\subsubsection{Exact expressions}

The inertial case is readily dealt with. The amplitudes are still given
by (\ref{AmplInertialLI}).
Provided condition (\ref{positivity cond}) is verified, ${\cal A}_+^{In}$
vanishes and ${\cal A}_-^{In} \neq 0$ for the wavenumbers allowed by the 
conservation of energy. If $v_\varphi < 1$ over some interval, 
the energy of the scalar quantum in the rest frame of the detector 
can become negative, thus opening the channel ${\cal A}_+^{In} \neq 0$ while
closing the other ${\cal A}_-^{In}=0$, which is again interpreted 
as the instability of the vacuum for frames such that
$\beta > v_\varphi$.

We now turn to the UA trajectory. 
We use the same notations as in sec. \ref{sec:ampLI} except for $k_\bot$
in eq. (\ref{amp1}) which should be replaced by its more general expression
$z=\sqrt{k_+ k_-}$, see eq. (\ref{kpm}).
The superluminal case is a repetition of the Lorentz invariant one because 
$\omega_k > k_z$ in every Lorentz frame. We get
\ba\label{AmplUAsuper}
   {\cal A}_{\pm,{\bf k}}^{UA} 
   = \frac{- i 2g }{\sqrt{2\omega_k (2\pi)^3} }\;
   \frac{e^{\mp\frac{\pi E}{2a}}}{a}  
   \left( \frac{k_+}{k_-}  \right)^{\pm i\frac{E}{2a}}
  K_{\pm i\frac{E}{a}}(\sqrt{k_+ k_-}) \, . \quad
\ea
The only essential difference is that $k_+ k_- = \omega_k^2 - k_z^2 \neq k_\bot^2$.
But since $K_{-\nu}(z)=K_\nu(z)$, the ratio of the probabilities 
is still a Boltzman factor
\ba \label{rapportBoltzmann}
   \frac{P_{+,{\bf k}}^{UA}}{P_{-,{\bf k}}^{UA}} = e^{-2\pi E/a} \, .
\ea

The description of the subluminal case with 
small values of the transverse wavenumber
is however different.
Now either $k_+$ or $k_-$ is negative. Let us say $k_- < 0$ for definiteness.
With the change of variables $y = x + \ln\sqrt{- \,k_-/k_+}$ we now have
\ba \label{amp2}
   {\cal A}_{\pm,{\bf k}}^{UA} \propto \left( - \frac{k_+}{k_-} \right)^{\pm iE/2} \, 
   \int_{-\infty}^{+\infty}\!\!dy \, e^{\pm i Ey - iz\cosh(y)} \, , \quad 
\ea 
where we note $z = \sqrt{-k_+ k_-}$.
This integral is defined as the analytic continuation of the modified Bessel function 
in terms of Hankel functions, 
$K_\nu(z) = -i\frac{\pi}{2} e^{-i\pi \nu/2} H_\nu\left(e^{i\pi/2}z\right)$.
\ba\label{AmplUAsub}
   {\cal A}_{\pm,{\bf k}}^{UA} 
   = \frac{- g \pi}{\sqrt{2\omega_k (2\pi)^3} }\;
   \frac{e^{\mp\frac{\pi E}{2a}}}{a}\left( -\frac{k_+}{k_-} \right)^{\pm i\frac{E}{2a}}
  H_{\pm i\frac{E}{a}}^{(2)}(z)  \, . \quad
\ea
Since $H_{-\nu}^{(2)}(z)=e^{-i\pi\nu}H_\nu^{(2)}(z)$, we now have
\ba \label{rapport unite}
   \frac{P_{+,{\bf k}}^{UA}}{P_{-,{\bf k}}^{UA}} = 1 \, . 
\ea
This result could have been almost anticipated from the 
response of the inertial detector since then we also noticed that 
both channels, excitation and desexcitation of the detector by emission
of a quantum, can occur. 
What is perhaps surprising about (\ref{rapport unite}) is that it is 
independent of the energy gap, as if the detector was coupled to a 
"reservoir" of infinite energy. 
We explain this by the fact that the vacuum defined in the preferred frame
is unstable in frames $\beta \geq \tanh(v_\varphi)$, hence along most of the 
UA trajectory. 
(The ratio (\ref{rapport unite}) is what one would 
obtain in a thermal bath at infinite temperature, but
one should not use this misleading analogy because the UA detectors does not 
react as in a thermal bath.)

\subsubsection{Steepest descent approximation}

Looking at the integral expressions (\ref{amp1}) and (\ref{amp2}), we see that 
they differ only by a shift $i\pi/2$ of the variable of integration, 
since $i\sinh(y) = \cosh(y +i\pi/2)$. 
A similar shift by $i\pi/2$ was found in the position of the simple poles 
of the Wightman function between the superluminal (\ref{poles v>1}) and 
subluminal DR (\ref{poles v<1}).
To better understand how these two analytic structures are related, 
we evaluate the previous integrals by the method of steepest descent.
\ba\label{steepest}
   I_{\eta} &=& \int\!\!\frac{dx}{\sqrt{2\pi}} \, e^{(E/a)f(x)} \, , \nonumber \\
   f(x) &=& i\eta x + \frac{i}{2}\left( k_- e^x - k_+ e^{-x} \right)
\ea
where $\eta=\pm 1$.
From now on we work in the units of $E$.
The results of the analysis, presented in appendix \ref{app:saddle}, are summarized
by the fig. \ref{fig12} and \ref{fig11}. The two cases they represent 
are respectively $\Delta > 0$ and $< 0$, where $\Delta = 1 + k_z^2 - \omega^2$.
The figures represent the real part of $f(x+iy)$, the dots are the 
saddle points $x_\pm$ (two in each case), and the curves are the path of 
stationary phase, i.e. the solutions of $\im(f[\gamma(\lambda)]) = \im(f(x_\pm))$.
Only in the case $\Delta < 0$ does this curve 
verify the additional condition
$\arg\left(f''(x_\pm)(x\vert_{\gamma}-x_\pm)^2 \right) = \pi$
so that it corresponds to the steepest descent path.

In more details, for $\Delta > 0$ (subluminal DR and slightly superluminal DR),
the steepest descent path does not exist. 
The curve of constant phase first climbs from an abyss before reaching the saddle point,
and leaves it to ascend a mountain. If it existed, the steepest descent path would 
instead continue progressing toward increasing $\im(x)$
in order to reach the abyss nearby. 
\begin{figure}[htp]
\hbox to\linewidth{\hss\resizebox{7.6cm}{6.7cm}{\includegraphics{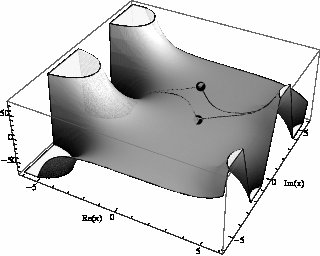}}\hss}
\caption{Region I and II of parameter space ($\omega, \, k_z$) corresponding to 
$\Delta < 0$. The real part of $f(x)$ as a function of $\re(x)$ and $\im(x)$.
The dots are the saddle points and the curves on the surface are the 
solutions of (\ref{stat phase}).
We took $\omega=1.5,k_z=2$ and $\eta=-1$. The figure for $\eta=+1$ is similar.\label{fig12}}
\end{figure}

For $\Delta < 0$ (superluminal DR),
the steepest descent path exists. 
It is the curve of stationary phase (\ref{solIII}) 
passing through the saddle point $x_+$.
\begin{figure}[h!]
\hbox to\linewidth{\hss\resizebox{7.6cm}{6.7cm}{\includegraphics{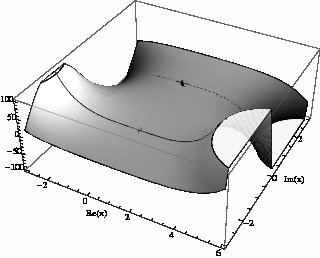}}\hss}
\caption{Region III of parameter space corresponding to $\Delta >0$. 
The real part of the phase $f(x)$ as a function of $\re(x)$ and $\im(x)$.
The dots are the saddle points and the curves on the surface are the steepest descent paths
(plain for $x_+$ and dashed for $x_-$).
We took $\omega=7,k_z=5$ and $\eta=-1$. The figure for $\eta=+1$ is similar.\label{fig11}}
\end{figure}
The crutial difference with the previous case is that
abysses on the side $\re(x) > 0$ face abysses on the opposite side,
while in parameter regions I and II their relative positions is shifted by $\pi/2$,
mountains thus facing abysses.
The real part of $f(x+iy)$ is indeed
\ba
\label{re f K-}
  &&k_- < 0 \,, \,  
  \re(f) = - 2\eta y + C \sinh\left( x + \frac{1}{2} \ln\frac{-k_-}{k_+} \right) \sin y
\, ,\nonumber \\
\label{re f K+}
   &&k_- > 0 \, ,\,
  \re(f) = - 2\eta y - C \cosh\left( x + \frac{1}{2} \ln\frac{k_-}{k_+} \right) \sin y
\, . \nonumber \\
\ea
The behaviour at large $x$ of the former expression, valid in regions I and II, 
depends on the sign of $x$. For $x > 0$, it diverges to $+\infty$
for $y \in \left] \pi,\, 2\pi \right[$, and for $x < 0$ it diverges to $+\infty$
for  $y \in \left] 0,\, \pi \right[$. Again, mountains are facing abysses.
The asymptotic behaviour of the second line of 
(\ref{re f K+}), relevant for region III, is independent of the 
sign of $x$ and mountains on one side thus face mountains on the other side.
We note finally that both the relative positions of the mountains and abysses and 
the angle of the tangent at the saddle point go hand in hand since $f(x)= i\eta x + f''(x)$, 
and it is $f''$ which determines both.

This establishes that the additional poles of the Wightman function 
(\ref{poles v>1}) and (\ref{poles v<1})
are in one-to-one correspondance with the positions of the maxima of 
$\re(f)$ (again, this was expected since transition amplitudes and Wightman
function are build from the modes).

We finish with the expression of the amplitude in region III evaluated on 
the steepest descent path 
\ba
  I_{\eta = -1} &=&
\frac{e^{- (E/a) (\pi -\arctan\sqrt{-\Delta}) - (E/a)\sqrt{-\Delta}}}{\sqrt{E/a}(\omega^2-1-k_z^2)^{1/4}} \left(\frac{k_+}{k_-}\right)^{i E/2a} 
\nonumber \\
  I_{\eta = 1} &=& e^{\pi E/a} I_{\eta = -1} \, .
\ea
Reminding that $K_\nu \sim \sqrt{\frac{\pi}{2z}} e^{-z}$ for $z\gg 1$, 
these expressions are indeed the $\omega\gg k_z$ limit of (\ref{AmplUAsuper}).

\subsection{Discussion}

We saw in sec. \ref{sec:withLI} that Lorentz symmetry assumes a double role:
it ensures the stability of the vacuum in any frame and 
endowes the UA trajectories with special properties
which gave in turn the stationarity and thermality of the Wightman function and
transition amplitudes and rates. 
In this section we first saw that stability is 
preserved only by superluminal DR (\ref{condpositivity}).
Second, stationarity and thermality are lost simultaneously. 
Third, the transitions are equal to the Lorentz invariant ones 
over a finite interval and deviate significantly from them outside
that interval. 
This interval is fixed by the value of the Lorentz factor (\ref{Gamma})
such that the detector probes the non trivial properties of 
the dispersion relation.
Quite remarkably, no matter how small the deviation from $\omega = k$ may be, once
$\Gamma$ is large enough the transition rates deviate appreciably from the 
Lorentz invariant ones.
With a linear DR (\ref{linDR}), fig. \ref{plotR} and \ref{plotRratio} show 
that the transition rates are exponentially close 
to the Lorentz invariant ones for $\vert \tau \vert \leq \rho$ and
exponentially small and equal for $\vert \tau \vert \geq \rho$. 
The transition between the two regimes is moreover sharp.
For more general DR, we showed in the appendix \ref{app:Taylor}
that the corrections are controled by the Lorentz factor 
times $E/M$ or $a/M$, see eqn. (\ref{rateTaylor}).

By Lorentz symmetry we have four equivalent ways to demonstrate 
the Unruh effect. This equivalence is lost in the present case. 
In particular the mathematical construction which consists in 
the definition of a quantum field theory in a Rindler wedge 
has no operational meaning (the Bogoliubov coefficients are 
not proportional to the transition amplitudes of detectors).
Note also that one should not interprete the
value of the ratio of the transition amplitudes for superluminal DR (\ref{rapportBoltzmann})
as the proof of the Unruh effect, because only inclusive probabilities are measurable. 
In relativistic theories we can use the ratio to characterize the effect because, by
Lorentz invariance of the measure $d^3k/\omega_k$, the amplitudes give the transition
rates (\ref{R+acc}). Without Lorentz invariance, 
the transition amplitudes sum up instead to non stationary transition rates.

In conclusion, we have seen that Lorentz invariance is both necessary and sufficient
to the Unruh effect which can be characterized in four equivalent ways, although 
only the transition rates of a detector are measurable quantities.
Without Lorentz symmetry, this equivalence is lost and the transition rates 
differ significantly from the Lorentz invariant ones after a proper time 
at which the Lorentz factor of the detector is high enough to probe the 
non trivial features of the dispersion relation ($v-1$ or $E/M$).
One expects however that this value is very high, simply because the scale at which
Lorentz invariance might occur is very high.  
It seems therefore unlikely that the Unruh effect could be used as a
practicle test of Lorentz invariance.

\section{Comparaison of the Unruh effect with Hawking radiation}
\label{sec:comparison}

Hawking radiation (HR) designates the property of the vacuum as defined by observers 
in free fall near the horizon to correspond to a thermal bath for distant 
observers.
The first four subsections contain review material. Their
presentation is however sketchy and the reader will find 
further details in 
references \cite{acoustic,Visser,Poisson,UnruhSchutzhold1} in particular. 
The comparison with the Unruh effect
is done in section \ref{sec:compare}.

To begin, it is worth recalling that
HR has little to do with gravity in the sense that
whether the metric is a solution of 
the Einstein's equations or not is irrelevant to the matter \cite{Visser}
(but field equations are necessary to establish the laws of black hole thermodynamics).
What does this mean exactly for a Schwarzschild black hole?
The metric outside the black hole in static form is
\ba \label{static metric}
  ds^2 &=& - f(r)dt^2 + \frac{dr^2}{f(r)} + r^2 d\Omega^2 \, , 
\nonumber \\
  f(r) &=& 1 - \frac{r_s}{r} \, .
\ea
The curvature is given by
$R_{ab} = 0$, $C^{abcd}C_{abcd} = 12 r_s^2/r^6$, and 
the surface gravity is $\kappa = GM/r_s^2$.  
Hawking's result uses only those properties of 
space-time which survive in the limit
\ba \label{limit}
  r_s = \frac{2GM}{c^2} = {\rm cte} \, , \quad
  M \to \infty  \, , \quad G \to 0
\ea
where gravity is decoupled while the geometry is fixed. 
Indeed in this limit 
the curvature and surface gravity, 
which are defined and calculated without recourse to the 
field equations, depend only on $r_s$ and are therefore invariant. 
Two additional quantities are important (the field equations
are required to show the following results). First 
the evaporation time $t_{\rm ev} \propto G^2 M^3/\hbar c^5 \to \infty$, 
that is backreaction vanishes, which shows that the limit of fixed background is self-consistent.
Second, the entropy $S = 4\pi GM^2/\hbar c \to \infty$, hence 
quantum field theory is in principle valid down to 
arbitrarily small scales. This is because to a region of size $r_s$ is associated 
a maximal density of state $e^{S}$ \cite{entropybound} in a dynamical theory of gravity. 
This statement is incompatible with a relativistic field theory which has an
infinite density of states by Lorentz invariance. 
Conversely if relativistic quantum field theories in curved space time are valid, 
the entropy of the black hole (which saturates Bekenstein's bound) must be infinite.

\subsection{Radially free falling frames}

Let us now proceed with the derivation of HR. 
As we said the state of the field is the vacuum as seen by geodesic observers.
Begin thus with the construction of 
a one-parameter family of coordinates $(\tau,\rho,\theta,\phi)$
attached to observers radially free-falling \cite{Poisson}.
This family is parameterized by the Lorentz factor $\gamma$ 
of the observers at infinity 
\ba
  \gamma^2 = \frac{1}{1 - v_{\infty}^2 }
 = \frac{1}{p} \, .
\ea
$\tau$ is their proper time 
and $\rho$ is the proper radial coordinate on the surfaces of constant $\tau$.
The radial velocity is given by
\ba
  v(\rho) = - \sqrt{1-pf[r(\rho)]} \, .
\ea
The sign in front of the square root is negative because we consider
infalling geodesics.
These coordinates are related to the Schwarzschild coordinates by 
$d\tau = \gamma \left( dt - \frac{v}{f}dr \right)$ and $d\rho = dr/\gamma$, or 
in integral form
\ba
  \rho = \frac{r}{\gamma} \, , \qquad \tau = \gamma \left( t + R(r) \right) 
\ea
with \footnote{We report a typo in eq. ($3.4$) of \cite{Poisson}. One should read
$\tau/\gamma=T=t + r\sqrt{1-pf} + ...$ instead of $T=t + r(1-pf) + ...$. 
It can be easely checked 
in taking the limit $\rho \to \infty$ that
the latter expression is incorrect.}
\ba 
  R(r) &=&- \,rv(r) + r_s \ln\left\vert\frac{r-r_s}{r_s}\right\vert 
\nonumber \\
       &&
 -\,\frac{(p-2)}{2 \sqrt{1-p}} \, 
        r_s \ln\left\vert p + 2 \frac{r}{r_s}\left(1 - p - \sqrt{1-p} \, v \right)\right\vert 
 \nonumber \\ 
 &&  - \,
       r_s\ln\left\vert pf(r) + 2 \left(1 - \frac{r}{r_s} v \right) \right\vert  
      + {\rm cte} \qquad
\ea
Only the near and far horizon limits are actually interesting
\ba \label{R limit hor}
  &&r\to r_s \, , \quad R(r) = r_s \ln\left(\frac{r-r_s}{r_s}\right) + C
    + O\left(\frac{r-r_s}{r_s}\right)  \nonumber 
\\
\label{R limit infty}
  &&r\to \infty \, , \quad R(r) = - v_\infty r + C' + O\left(\frac{r_s}{r}\right) \, .
\ea
Near the horizon the $R(r) \sim r_*(r) = r + r_s \ln(r/r_s - 1)$, the tortoise coordinate
defined by $dr_* = dr/f(r)$, so $\tau \to \gamma( t + r_*)$ where $t+r_*$ is 
the advanced Eddington-Finkelstein null coordinate. 
Far from the horizon, $\tau \to \gamma (t - v_\infty r)$ is naturally 
the proper time of observers boosted with the static observers w.r.t. the black hole.
The line element
\ba \label{PG}
  ds^2 &=&  - \left( 1- v^2 \right) d\tau^2 - 
   2 v d\tau d\rho + d\rho^2 + r^2 d\Omega^2 \qquad
\ea
is stationary and rotationaly invariant.
The radial component of the shift vector is $-v(\rho)$.
The Schwarzschild radius $r=r_s$ corresponds to
$v(\rho_s)=-1$ and near the horizon the expansion
\ba \label{vhor}
  v = - 1 + x + O(x^2) \, , \qquad x = \kappa_\gamma (\rho - \rho_s)  
\ea
will be used, where 
\ba
  \kappa_\gamma = \frac{1}{2r_s \gamma}
\ea
is the surface gravity seen by these observers at infinity.

\subsection{Partial wave decomposition of the field equation}

We now pick one of these free fall frames and define the 
modified dispersion relation from the quadratic action
\footnote{It is customary to choose the frame at rest at infinity as 
the preferred frame. This is probably an implicit simplifying assumption, or
the natural choice in the idealized model of an isolated black hole. 
There is otherwise no reason to opt for this frame, to which one could
prefer for instance the rest frame of the cosmic microwave background.
We shall therefore not follow the custum. Besides, this does not introduce 
any complication because only the asymptotic behaviours 
(\ref{R limit hor}) of the free fall 
coordinates and modes (\ref{def phi hr}) and (\ref{phi hr}) are relevant, 
and they are universal up to a trivial factor of $\gamma$.}
\ba \label{action}
  S &=& - \frac{1}{2}\int\!\!d\tau \int\!\!d^3x \, \sqrt{q} \left\{ 
   (n^a \partial_a \phi)^2 
 \right. 
\nonumber \\
&& \qquad \left.
+ \,\, q^{ab}\partial_a \phi  \partial_b \phi 
   + \phi F(\Delta) \phi
 \right\}
\ea
written in covariant form with the help of the $1+3$-decomposition of the metric:
$n=(1,v,0,0)$ is the unit vector normal to the surfaces of constant $\tau$,
$q^{ab} = g^{ab} + n^a n^b$ is the  induced (contravariant) metric tensor on 
these surfaces, and 
$\Delta = q^{-1/2} \partial_a\left( \sqrt{q}q^{ab} \partial_b \right)$ 
the corresponding Laplacian.
In the following two subsections we review in turn the Lorentz invariant case $F=0$, 
and the changes introduced by $F \neq 0$.

The Klein-Gordon scalar product
\ba \label{KG}
  ( \phi,\, \psi ) = i \int\!\!d^3x \, \sqrt{q} \left\{ \phi^*(n^a\partial_a\psi) - 
  \psi (n^a\partial_a\phi)^*  \right\}
\ea 
is conserved on the solutions of the field equation. This equation 
\ba \label{cov eom}
  \partial_b\left(\sqrt{q} \, n^b n^a \partial_a\phi  \right) = \Delta \phi + F(\Delta) \phi
\ea
is separable for the ansatz
 \ba \label{sep var}
  \phi_{\omega l m} = 
   \frac{e^{-i\omega \tau}}{r} Y_{lm}(\theta,\phi) \, \varphi_{\omega l}(\rho)
\ea
where the radial functions $\varphi_{\omega l}$ are solutions of 
\ba \label{general EOM}
  &&(v^2 -1)\varphi'' + 2v(v'-i\omega)\varphi'- \omega(\omega+iv')\varphi  
  \nonumber \\
  &&\quad 
 = \,\, \left[ {\cal V}_l +  F(D) \right] \varphi \, .
\ea
The prime stands for the derivative with respect to $\rho$.
The auxillary functions appearing in this equation are the transverse momentum
\ba \label{k_l}
  k_l^2 = \frac{l(l+1)}{r^2} \, \, , \qquad
\ea
and the effective potential
${\cal V}_l = k_l^2 - 2 vv'\frac{r'}{r}$. We introduced the 
second order derivative operator $D$ defined by
\ba 
 \Delta \phi_{\omega l m} \equiv r^{-1} Y_{lm} \left( D \varphi_{\omega l} \right)
\ea
Its explicite expression is
$D \varphi_{\omega l} = \left( \partial_\rho^2 - k_l^2 \right) \varphi_{\omega l}$.

\subsection{HR with Lorentz invariance}
\label{sec:HR with LI}

In this section, $F = 0$.
Equation (\ref{general EOM}) has the particular property that the
coefficient of $\partial_\rho^2 \varphi$ vanishes at the horizon.
It must therefore be solved separately on each side of the horizon. 
We thus expect two classes of solutions, the ones that are
singular at the horizon and the ones which are regular.
They will describe respectively radiation at infinity and the free fall vacuum.
HR is established by showing that these two sets of modes are related by a
Bogoliubov transformation similar to (\ref{asymp phiU})-(\ref{alpha}).

This is most simply done by solving the field equation in the eikonal approximation
\ba
  &&\phi = \frac{1}{r}\, {\cal A}(\tau,\rho,\theta) \, e^{iS(\tau,\rho,\theta)} 
\, , \quad 
   g^{ab}\partial_a S \partial_b S = 0\qquad 
\ea
where ${\cal A}$ is a slowly varying function compared to the Hamilton-Jacobi action $S$. 
By separation of variables, 
$S = - \omega \tau + \sqrt{l(l+1)} \theta + \int^\rho\!\!d\rho' k_\omega(\rho')$ where
$k_\omega$ are the solutions of the quadratic equation
\ba \label{eqn eikonal}
  \left(\omega - v(\rho) k_\omega \right)^2 = k_\omega^2 + k_l^2 \, .
\ea
with $k_l$ defined at eq. (\ref{k_l}).
The two roots are
\ba \label{k eikonal}
  k_\pm(\omega,\rho) = \frac{1}{1-v^2(\rho)}\, 
  \left( -\omega v \pm \sqrt{\omega^2 + (v^2-1)k_l^2} \right) \quad
\ea 
Far from the horizon $v \to v_\infty$, $k_l \to 0$, and 
$k_\pm \to \omega/(v_\infty \pm 1)$. 
The solutions $k_-(\omega) < 0$ are propagating towards the black hole 
(thus describe infalling radiation)
\ba 
  \rho \to \infty \, , \quad 
  \phi^{\rm inf} &\to& \frac{{\cal A}}{r}
  \exp\left\{ -i\omega \left( \tau + \frac{\rho}{1-v_\infty} \right) \right\} \, 
\nonumber \\
 \qquad  &=& \frac{{\cal A}}{r} e^{ -i\gamma \omega \left( t +  r\right) } \, .
\ea
To get the second expression we substituted eq. (\ref{R limit infty}).
The solutions $k_+(\omega) > 0$ 
propagate away from the black hole and describe the HR emerging at infinity
\ba \label{def phi hr}
  \rho \to \infty \, , \quad 
  \phi^{\rm hr} &\to& \frac{{\cal A}}{r}
  \exp\left\{ -i\omega \left( \tau - \frac{\rho}{1+v_\infty} \right) \right\} \, 
  \nonumber \\
  \qquad &=& \frac{{\cal A}}{r} e^{-i\gamma \omega(t-r)} \, .
\ea
Near the horizon, the infalling solution becomes 
$k_- \to - \frac{\omega}{2}  + \frac{k_l^2}{2\omega} + O(x)$ and 
\ba  \label{phi inf}
  \rho \to \rho_s^+ \, , \quad 
  &&\phi^{\rm inf} \to  e^{-i\omega(\tau + \rho/2)} \, 
\ea
and does not present any particular interest. The state of these modes is
assumed to be the vacuum.
The other solution $k_+  \sim \frac{\omega}{x}$ is singular at the horizon 
and describes two types of modes with support on either side of the horizon
\sba \label{phi hr}
  \rho \to \rho_s^+ \, , \quad 
  && k_+ \to \frac{\omega}{x} - \frac{1}{2} \left(\omega + \frac{k_l^2}{\omega} \right) 
    + O(x)   \, , \\
  &&\phi^{\rm hr} \to 
  \theta(x) \, 
  \exp\left\{ -i\omega \left( \tau  
   - \frac{1}{\kappa_\gamma} \ln x  \right) \right\} 
\nonumber \\
  &&\qquad = \theta(r-r_s) \, e^{-i\gamma \omega (t-r_*)} \, , \\
  &&\phi^{\rm ptn \, *} \to 
  \theta(-x) \, 
  \exp\left\{ -i\omega \left( \tau  
   - \frac{1}{\kappa_\gamma} \ln (-x)  \right) \right\} 
\nonumber \\
  &&\qquad = \theta(r_s-r) \, e^{-i\gamma \omega (t-r_*)} 
\sea
where we have omitted terms $O(1)$ in the phase.
The modes $\phi^{\rm hr}$ describe Hawking radiation. Note that the asymptotic forms 
(\ref{phi hr}b) and (\ref{def phi hr}) can be written $e^{-i\gamma \omega u}$
in terms of the advanced null coordinate $u = t- r_*$.
The modes $\phi^{\rm ptn \, *}$ are trapped inside the horizon and 
describe the partners of the quanta of Hawking radiation. They are complex conjugated 
so that they have a positive norm.

This completes the description of the solutions relevant for the quantization
by observers at infinity.
Freely falling observers on the other hand 
are assumed to experience the vacuum as they cross the 
horizon. One defines the free fall vacuum in a similar way to the Unruh vacuum in 
sec. \ref{sec:stateLI}. One constructs a basis of modes $\phi^U$ 
regular across the horizon and with the 
following properties:\\
i) they are eigenmodes of the Killing vector $\partial_\tau$, 
\ba \label{sep var}
  \phi^U_{\omega l m} = 
   \frac{e^{-i\omega \tau}}{r} Y_{lm}(\theta,\phi) \, \varphi^U_{\omega l}(\rho)
\ea
ii) they are eigenfunction of the Lie derivative with respect to 
the unit vector $n$ orthogonal to the surfaces of constant $\tau$
\ba \label{ff modes}
  {\cal L}_{n}\, \phi^U_{\omega l m}
  = n^a \partial_a \phi^U_{\omega l m}
   = -i\Omega(\rho) \, \phi^U_{\omega l m}  \, .
\ea
with $\Omega > 0$, so that they have positive norm (\ref{KG}).
One notices that near the horizon we have the identity 
\ba \label{redshift}
   {\cal L}_{n}\,  e^{-i\gamma \omega (t-r_*)} 
   = -i \frac{\omega}{\kappa_\gamma(\rho - \rho_s)}\,  e^{-i\gamma \omega (t-r_*)}
\ea
Hence the solutions $e^{-i\gamma \omega (t-r_*)}$ have positive (negative) 
free fall frequency outside (inside) the horizon. 
Similarly, the solutions $e^{+i\gamma \omega (t-r_*)}$ have positive free fall  
frequency inside the horizon. Reminding that 
$e^{i\omega r_*} = x^{i\omega/\kappa_\gamma}$, 
one infers that near the horizon
\ba
  \phi^U_{\omega l m} \sim \alpha_{\omega l}e^{-i\omega t}\,
  \vert x \vert^{i\omega/\kappa_\gamma} 
  \left\{ \theta(x) + \frac{\beta_{\omega l}}{\alpha_{\omega l}} \theta(-x)\right\} 
\ea
(we suppressed  the term $Y_{lm}$ for a better lisibility).
The unique analytic continuation of $x^{i\omega/\kappa_\gamma}$
which is bounded in the domain
$\left\{ \im(t) < 0 , \, \im(r_*) \leq -\im(t) \right\}$ is
\ba \label{Unruhmode}
   \varphi^{U}_{\omega l } 
   &=&  \alpha_{\omega l} \left(x +i\epsilon\right)^{i\omega/\kappa_\gamma} 
\ea
The coefficient $\alpha$ is fixed by the normalization.
With the branch cut of the logarithm along the negative real axis, the r.h.s. 
evaluates to the sum
\ba \label{HRLI}
  \phi^{U}_{\omega lm} = \alpha_{\omega l} \phi^{\rm hr}(\rho) + 
  \beta_{\omega l}\phi^{\rm ptn \, *}(\rho)
\ea
with the ratio of the Bogoliubov coefficients
\ba \label{BogolHR}
  \left| \frac{\beta_{\omega l}}{ \alpha_{\omega l}}\right|^2 
  = e^{-2\pi \omega/\kappa_\gamma} \, .
\ea

\subsection{HR without Lorentz invariance}
\label{sec:HRLV}

There are three key elements in the derivation of (\ref{HRLI}).
One is the universal behaviour of the coordinate $\tau(t,r)$ near the horizon.
Modifying the dispersion relation does not change this.
The second is the logarithmic dependence of the modes 
near the horizon as a function of $x$. The third is the branch cut associated 
with the analytic extension of this logarithm. It was introduced
when we chose the state to be the free fall vacuum.
Let us now examine the changes caused by $F \neq 0$.
The following discussion is mainly qualitative.

The origin of the logarithm is the "kinetic" term of eq. (\ref{general EOM}),
i.e. the differential operator on the l.h.s of this equation.   
We can factorize the latter into an infalling 
and an outgoing part $\partial_{\rm inf} \partial_{\rm hr} \varphi$ where 
\sba \label{factor kin}
  \partial_{\rm inf} &=& \partial_\tau +  \partial_\rho v - \partial_\rho \to 
  \partial_\tau - 2 \partial_x 
\\
  \partial_{\rm hr} &=& \partial_\tau + v \partial_\rho + \partial_\rho \to  
 \partial_\tau + \kappa x \partial_x
\sea
The solution of the former is $\phi^{\rm inf}$
of eq. (\ref{phi inf}), and the solution of the second 
describes Hawking radiation (\ref{phi hr}). 
Since this operator is the difference of the l.h.s. of (\ref{cov eom}) 
with the Laplacian $\Delta \phi$, 
whenever one replaces in the action 
$\phi \Delta \phi$ by $\phi (\Delta + F(\Delta)) \phi$ as in (\ref{action}),
this kinetic operator is preserved. So we also expect to find solutions with a 
branch cut in that case.

Indeed, taking the Fourier transform w.r.t. $x$ of the limiting expressions of 
(\ref{factor kin}) near the horizon gives
\ba \label{TFeom}
  i(\omega + 2k) \left[\kappa \partial_k + (i\omega + \kappa ) \right] \tilde \varphi = 
  \left[ {\cal V}_l +  {\cal F}(k^2, k_l^2) \right]  \tilde \varphi \qquad 
\ea
where ${\cal V}_l$ is a constant near the horizon.
Note that the function ${\cal F}$ 
differs from $F$ of the dispersion relation by terms containing derivatives of $k_l^2$,
possibly coupled to derivatives of $\varphi$.
For instance, a term $D^2 \varphi$ gives
$\left[ k^4 + 2 k_l^2 k^2 +
 \left( k_l^4 - \partial_\rho^2 k_l^2 \right) \right] \tilde \varphi$.
Similarly, a term $D^n \varphi$ produces a homogeneous polynome of order $2n$ 
multiplying $\tilde \varphi$.
Only for the $s$-wave (or in $1+1$ dimensional models) 
do we have ${\cal F}(k^2, k_0^2) = F(k^2)$.
These terms could affect significantly the grey body factor.
Let us now consider the left hand side of (\ref{TFeom}).
In relativistic theories, we know that first 
black holes are black bodies, so the typical frequency of Hawking radiation
is given by the temperature, i.e. $\omega = O(\kappa)$, and second, 
the physically interesting region is $\kappa_\gamma x \ll 1$.
In other words HR corresponds to low frequencies and high wavenumbers $k \gg \omega$. 
We thus replace $\omega + 2 k$ on the l.h.s. of (\ref{TFeom}) by $2k$ 
and solve the equation with the ansatz 
$\tilde \varphi =\varphi_0(k) \,\chi(k)$, 
with $\varphi_0 = \theta(k) k^{-i\omega/\kappa_\gamma -1}$ solution of 
$\left[\kappa \partial_k + (i\omega + \kappa ) \right] \varphi_0 = 0$
defining the free fall vacuum 
(since $\varphi_0(x) \propto (x + i\epsilon)^{i\omega/\kappa_\gamma}$), 
and $\chi$ solution of $d \ln \chi = -i {\cal F}/2\kappa k^2$. 
This approximation, also adopted in \cite{UnruhSchutzhold1}, amounts to 
neglect the coupling between the outgoing and infalling solutions (the latter 
corresponding to the root $\omega + 2k = 0$ as we know from the WKB solution).
This coupling can indeed be argued to be innocuous \cite{UnruhSchutzhold2}.
Under certain assumptions, e.g. analyticity of $F$, 
the inverse Fourier transform can be estimated in 
the steepest descent approximation, in which case HR is found, 
see \cite{UnruhSchutzhold1} and \cite{Corley} for more details.
Its origin is clearly identified as the branch cut of $\varphi_0$.

This leaves the question of the state. 
At least one condition seems necessary so we can assume that the field is in the 
free fall vacuum, 
namely that the evolution of the modes is adiabatic \cite{acoustic,UnruhSchutzhold1}.
This places certain constrains on the dispersion relation.
For instance, if $F$ is polynomial of order $2n$, the modified dispersion
relation $(\omega - v k)^2 = k^2 + k_l^2 + {\cal F}(k^2,k_l^2)$ 
posesses $2n$ solutions amongst which
$2p \geq 2$ are real and $2(n-p)$ are complex conjuguate.
In that case a necessary condition for adiabaticity 
is the absence of level crossing between the real roots. 
This can happen via a kind of seesaw mechanism 
if the modifications of the 
dispersion relation are characterized by a very high scale $M \gg \omega \sim \kappa$.
It requires some care to analyse the contribution of the complex roots, 
but again they should not affect the low energy part of the spectrum, 
adiabaticity implying their decoupling from 
the high energy modes. A detailed analysis of these complex roots 
for the DR $k^2 \pm k^2 / M^2$ confirms this
qualitative argument \cite{Machet1}.

The fundamental part played by adiabaticity should not be surprising if one recalls that 
modified actions such as (\ref{action}) 
describe an effective field theory,  
and in a non trivial background, both scale separation and adiabaticity are 
necessary to validate this framework \cite{adiabCST}.

\subsection{Comparison}
\label{sec:compare}

Let us finally return to the question of the relationship between the 
Unruh effect and Hawking radiation.
HR is habitually proved by finding the ratio of the 
Bogoliubov coefficients as in eq. (\ref{BogolHR}) and is 
therefore determined by the solutions of the field equation.
One set of modes (and therefore the corresponding state) 
are regular across the horizon (Minkowski/Unruh-like), while
the other set (Rindler/outgoing) have a logarithmic singularity (in $V=t+z$ or $x$). 

The essential difference is the role played by the horizon in the 
dynamics of the field.
There are actually two notions of "horizon" that should be distinguished.
One is the surface of infinite redshift associated with the observers, that is 
the surface $v=-1$ for observers far from a black holes, as examplified by 
(\ref{phi hr}) and (\ref{redshift}), 
and the null planes ${t=\pm z}$ for the uniformaly
accelerated observers.
The other is the locus of the logarithmic singularity of the modes.
Without Lorentz invariance, these two notions still 
coincide for a black hole, but they differ in flat space.
Indeed, whether the dispersion relation is relativistic or not, 
the field equation (\ref{general EOM}), or (\ref{factor kin}b), 
is singular at $v=-1$.
In contrast, the horizon of uniformaly accelerated observers is not the locus 
of singularity of the (pseudo-)Rindler modes, which
is $v_0 t=\pm z$ for linear dispersion relations $\omega_k = v_0 k$, 
and which is not defined
for general dispersion relations because of the non linear mixing between
Rindler coordinates.
This explains also why the dispersion relations (\ref{SO(1,1)}) with a 
$SO(1,1)$ symmetry mimic better the black hole context, because in that 
case the modes are still singular on the observer's horizon.

As any phenomenon of pair creation from an unstable ground state, 
Hawking radiation is characterized by a branch cut, namely (\ref{Unruhmode}),
or more generally $k^{-i\omega/\kappa_\gamma}$ from (\ref{TFeom}). 
This branch cut is robust because
the higher derivative terms do not mix with the operator
on the left hand side of (\ref{general EOM}) or its Fourier transform (\ref{TFeom}).
We already gave a necessary condition for this: that the gradients term in the 
action be replaced by $\Delta + F(\Delta)$.
There is a second condition that we did not mention so far, although it should 
be quite obvious. All the results of the 
previous sections depend on the fact that the observers at infinity 
and the observers near the horizon belong to the same referential,
that is they are all freely falling observers 
characterized by the same Lorentz factor $\gamma$.
This should be contrasted with the Unruh effect where it is necessary to 
boost a detector continuously (at a constant acceleration).
In the case of HR on the other hand, the redshift between free fall observers near and
far from the horizon is purely gravitational, that is caused by the 
curvature.
An observer equiped with a two-level detector and freely falling but with a 
different Lorentz factor $\gamma'$ would observe similar phenomena as
the ones described in the first part of the paper. For instance, 
if the dispertion relation is subluminal, it would perceive the 
free fall vacuum of the other observers as unstable if it is 
sufficiently boosted w.r.t. them.

One sometimes invokes the equivalence principle as the reason for the analogy 
between the Unruh effect and Hawking radiation. This is clearly not correct, for 
otherwise we would not expect to find a Hawking-like radiation of phonons 
in a variety of condensed matter systems (dumb holes) with an acoustic horizon, since 
this prediction does not require either
the Einstein's equations, nor even Lorentz symmetry.  
What the equivalence principle does imply however is that, if the preferred 
frame is not the one with $\gamma = 1$ (and there is no reason
why it should be, see footnote $2$), a static observer at 
fixed radial distance from the black hole should record 
transition rates similar to those of sec. \ref{sec:RLV}.

In brief, the prediction of Hawking radiation rests on the fact that 
the field equations in a black hole metric are singular on a surface 
which coincides with the horizon of asymptotic observers. As long as modified
dispersion relations do not alter this property of the field equation  
(and provided the evolution of the state is adiabatic), Hawking radiation 
is expected to be robust (in the preferred frame).
By contrast in $3+1$ Minkowski space, the locus where the  
the Rindler modes are singular coincides with the horizon of uniformaly accelerated
observers only if Lorentz invariance is assumed.

\begin{appendix}

\section{Bogoliubov coefficients}
\label{app:Bogol}

Since Rindler and Minkowski modes are solutions of a linear equation
and form complete families, they are related 
by linear transformations
\ba \label{Bogol}
  \varphi_\lambda^R &=& \int_{-\infty}^{+\infty}\!\!dk_z \,
  \left( \alpha_{\lambda k_z}^R \varphi_{k_z}^M + 
  \beta_{\lambda k_z}^R \varphi_{k_z}^{M \,*}  \right) 
\nonumber \\
  \varphi_\lambda^{L} &=& \int_{-\infty}^{+\infty}\!\!dk_z \,
  \left( \alpha_{\lambda k_z}^L \varphi_{k_z}^M + 
  \beta_{\lambda k_z}^L \varphi_{k_z}^{M \,*}  \right) 
\ea
From the property (\ref{phiL}) one deduces
\ba \label{RtoL}
  \alpha_{\lambda k_z}^L = \alpha_{\lambda\, -k_z}^{R} 
 \, , \qquad 
  \beta_{\lambda k_z}^L = \beta_{\lambda\, -k_z}^{R} \,  .
\ea
The Bogoliubov coefficients $\alpha$ and $\beta$ are given by the Klein-Gordon 
products
$\alpha^R = \left( \varphi^M,\, \varphi^R \right)$ and 
$\beta^R = \left( \varphi^{M\, *},\, \varphi^R \right)$.
Let us present the calculation of $\alpha^R_{\lambda k_z}$.
Since the Klein-Gordon product is independ of the time argument of solutions of 
the wave equation, we evaluate it at on the horizon ${\cal H}^+$.
After an integration by parts we have
\ba
  \alpha^R_{\lambda k_z} &=& i2 \int_0^{\infty}\!\!dV \, \varphi_\omega^{M *} \,
  \partial_V \varphi_\lambda^R + {\rm bnd}
\nonumber \\
  &=& 2\lambda {\cal N}_\omega {\cal N}_\lambda \int_0^{\infty}\!\!\frac{dV}{V} \,
  e^{i \omega_- V/2} \, V^{-i\lambda} + {\rm bnd} \, . \qquad
\ea
We note ${\cal N}$ the normalization constants of the modes. The boundary term is
$\left[ \varphi_\omega^{M *}\varphi_\lambda^R \right]_{V=0}^{V=\infty}$.
To obtain the second line we used $\varphi^M = {\cal N}_\omega e^{-i\omega_- V/2}$ on 
${\cal H}^+ = \left\{ U=0 ,\, V>0 \right\}$ with $k_- = \omega_k -k_z$ 
(see eq. (\ref{kpm})), and the 
second term in the asymptotic expansion (\ref{asymptphiR}) (we assume again
narrow wave packets in $\lambda$).
The integral can be written as the limiting value of Euler's $\Gamma$ function
\ba
 \int_0^{\infty}\!\!\frac{dt}{t} \, t^z e^{-kt} = k^{-z} \Gamma(z) \, , 
 \quad \re(z) > 0 \, , \,\,\re(k) > 0 \, . \quad 
\ea
with both $\re(z)$ and $\re(k) \to 0$.
With these regularizations, the boundary term vanishes and 
one finds finally 
\ba
  \alpha^{R}_{\lambda,k_z} &=& 
  \frac{e^{\pi \lambda}}{\left[ 4\pi \omega_k \sinh(\pi \lambda)\right]^{1/2}}
  \left( \frac{k_+}{k_-} \right)^{-i\lambda}
\nonumber \\
  \beta^{R}_{\lambda,k_z} &=& - \, e^{-\pi \lambda}  \, \alpha^{R}_{\lambda,k_z}
\ea
Comparing these expressions with (\ref{ampUALI}) yields (\ref{A proto alpha}). 

From the definition (\ref{Bogol}) and the mode expansions (\ref{Phi}) 
and (\ref{PhiRindler}) one obtains the relation between the Minkowski and Rindler
creation and annihilators from which eq. (\ref{interp1}) follows
(one could also obtain them from (\ref{Bogol operateurs}) and eq. (\ref{intermed}) below)

The coefficients $A$ and $B$ in (\ref{CL mink}) are then easely calculated.
Let us derive the coefficients $A_{\Omega k_z}$.
Substitute the expansion (\ref{Bogol}) in the r.h.s. of eq. (\ref{phiU}a) and 
regroup the terms multiplying $\varphi^M$ and $\varphi^{M\,*}$
\ba \label{intermed}
  \varphi^U_\Omega &=& \int_{-\infty}^{+\infty}\!\!dk_z \, 
  \left( \alpha_\Omega \alpha^R_{\Omega k_z} 
  +  \beta_\Omega \beta^{R\,*}_{\Omega\, -k_z}  \right)
  \varphi_{k_z}^M
\nonumber \\
  &&\qquad +\,  \left( \alpha_\Omega \beta^R_{\Omega k_z} 
   +  \beta_\Omega \alpha^{R\, *}_{\Omega\, -k_z}  \right)
  \varphi_{k_z}^{M \,*}
\ea
We used the property (\ref{RtoL}). 
The terms $\varphi_{k_z}^{M \,*}$ are absent from (\ref{CL mink}). 
Since the Minkowski modes form an 
orthonormal basis, each term in the brackets multiplying $\varphi_{k_z}^{M \,*}$ must
vanish and therefore
\ba \label{egality Bogol}
  \frac{\beta^R_{\Omega k_z}}{\alpha^{R \,*}_{\Omega \, -k_z}} 
   = - \frac{\beta_\Omega}{\alpha_\Omega} \, .
\ea
Finally substituting these expressions into the first line of (\ref{intermed}) one gets
with the help of the unitarity relation 
$\vert \alpha_\Omega \vert^2 - \vert \beta_\Omega \vert^2 = 1$
\ba
  A_{\Omega k_z} = \frac{\alpha^R_{\Omega k_z}}{\alpha_\Omega} \, .
\ea
Similarly one obtains $B_{\Omega k_z} = \beta^R_{\Omega k_z} / \beta_\Omega$.

It is a good check to proove with these expressions the orthogonality and 
completeness of the Unruh modes.
For instance with $\Omega$ and $\Omega' > 0$,
\ba
  \left( \varphi^U_\Omega ,\, \varphi^U_{\Omega'} \right) 
  &=& \int_{-\infty}^{\infty}\!\!dk_z \,  A_{\Omega k_z}^* A_{\Omega' k_z} 
\nonumber \\
  &=& \frac{1}{\alpha_\Omega  \alpha_{\Omega'}} \frac{e^{\pi(\Omega+\Omega')/2}}{4\pi 
  \left( \sinh \pi \Omega  \sinh \pi \Omega' \right)^{1/2}}
\nonumber \\
  && \, \times \int_{-\infty}^{+\infty}\!\!\frac{dk_z}{\omega_k} \, 
  \left(\frac{\omega_k + k_z}{\omega_k - k_z}  \right)^{i(\Omega - \Omega')}
\ea
In the first line we used the orthogonality of Minkowski modes (\ref{Minkmodes}).
The integral becomes trivial after the change of variable
$x = \ln\left(\frac{\omega_k + k_z}{\omega_k - k_z}  \right)$
and the factors combines to give
\ba
  \left( \varphi^U_\Omega ,\, \varphi^U_{\Omega'} \right) = \delta(\Omega - \Omega') \, .
\ea

\section{Taylor expansions}
\label{app:Taylor}

We consider dispersion relations admitting a Taylor expansion
\ba \label{Taylorv}
  v_\varphi(k) = \sum_{n=0}\frac{\alpha_n}{n!} \left(\frac{k}{M}\right)^{n} 
               = v  + f(k)\, . 
\ea
We assume its convergence radius infinite.
We renamed $\alpha_0 = v$ to be in keeping with the notation (\ref{linDR}).
We substitute (\ref{Taylorv}) into the integral expression (\ref{Wdk})
of the Wightman function and expand 
in terms of the $\alpha_{n \geq 1}$, keeping the first term $v$ in the phase 
\begin{widetext}
\begin{eqnarray} \label{intermed1}
  W &=& \frac{-i}{8\pi^2 r} \int_0^{\infty}\!\!\frac{dk}{v + f} 
  \left( e^{-ik(v t + r)} - e^{-ik(v t - r)}\right) e^{-ikt f(k)} e^{-\epsilon k}
\nonumber \\
    &=& \frac{-i}{8\pi^2 v r} \int_0^{\infty}\!\!dk \,
    \left( e^{-ik(v t + r)} - e^{-ik(v t - r)}\right) e^{-\epsilon k}  \, 
    \sum_{n=0}^{\infty} \beta_n(t) \left(\frac{k}{M}\right)^{n} \, .
\end{eqnarray}
\end{widetext}
Provided that $v \geq 1$, which we will assume in the following, 
we can replace $e^{-\epsilon k}$ by $t-i\epsilon$, see the discussion 
of sec. \ref{sec:W LV}. 
Note that even if the DR is polynomial, the Taylor expansion 
of the two-point function has an infinite number of terms.
We want to integrate this series term-by-term over $k$.
The exchange of the series and integral is permitted  
if and only if the series has an infinite convergence radius. 
Otherwise the result of the integration gives an asymptotic expansion of the Wightman
function.
We were not able to determine the convergence radius of the series 
under the integral sign, but the form of the result indicates that 
we are doing in fact an asymptotic expansion. 
Carrying out this integration we obtain indeed 
\ba \label{TaylorW}
  W(x_1,x_2) &=& -\inv{4\pi^2} \frac{1}{s^2} \left\{ w_0 + i \frac{w_1}{M s}
  - \frac{w_2}{(Ms)^2} + ... \right\} \qquad
\ea
We note $s^2 = -(x_1 - x_2)^2$ the invariant distance, and we give the 
expression of the coefficients in terms of the Lorentz factor
$\Gamma= (x_1^0 - x_2^0)/s$
\begin{widetext}
\ba
 w_0 &=& \inv{v(1+\Gamma^2(v^2-1))}
\, , \qquad
 w_1 = \frac{8 \Gamma^3v}{(1+\Gamma^2(v^2-1))^3}\, \alpha_1
\, , \\
 w_2 &=& \left\{ \alpha_2 \, v \left[ (\Gamma^2-1)^3
  -11 v^2 \Gamma^2(\Gamma^2-1)^2
  -5v^4 \Gamma^4(\Gamma^2-1) + 15v^6\Gamma^6\right]   \right.
\nonumber \\
  && \hspace*{-0.5cm} \left. 
  - 2 \alpha_1^2 \left[ (\Gamma^2-1)^3
-5v^2\Gamma^2(\Gamma^2-1)^2
+55v^4\Gamma^4(\Gamma^2-1)
+45v^6\Gamma^6\right] \right\} \inv{v^3(1+\Gamma^2(v^2-1))^5}
\ea
To simplify the discussion we take $v=1$ from now on.
On inertial trajectories with Lorentz factor 
$\gamma$, the r.h.s. of (\ref{TaylorW}) is given by
\begin{eqnarray}\label{TaylorWIn}
  {\cal W}_{In}(\tau) = -\inv{4\pi^2\tau^2} \left\{ 1
    + i\frac{8\gamma^3\alpha_1}{M\tau}
    + \frac{24}{(M\tau)^2} \left[ \alpha_2\gamma^4 \left( 1 + {\cal O}(\gamma^{-2}) \right)
    - 8 \alpha_1^2 \gamma^6 \left(1+ {\cal O}(\gamma^{-2}) \right) \right]
    + ... \right\} \, , 
\end{eqnarray}
\end{widetext}
and $\tau$ stands for $\tau-i\epsilon$.
The series on the r.h.s. of (\ref{TaylorWIn}) 
has an essential singularity at $\tau =0$, so it cannot
be equal to the Wightman function which is a tempered distribution. 
We thus conjecture that for generic dispersion relations, 
the integration term-by-term of (\ref{intermed1})
gives an asymptotic expansion of the Wightman function.
The expansions (\ref{TaylorW}) and (\ref{TaylorWIn}) then make sense only 
for $M s \gg 1$ and they must be truncated. 
(The order of truncation chosen to minimize the error depends on the 
detailled behaviour of the $w_n$, which we do not know.)

Now, we want to substitute this expansion into (\ref{Rpm})
and integrate term-by-term in order to get an expansion in 
powers of $E/M$.  
To be consistent with $M\tau \gg 1$, we are limited to 
energies
\ba \label{condasympt}
  E \ll M \, .
\ea
Luckily this is compatible with the condition 
$E\tau \gg  1$ (the Golden rule limit), necessary for the notion of a transition 
rate to be meaningful, and for the expression (\ref{Rpm}) to be 
valid. The result is
\begin{widetext}
\ba
  R_-^{In} = \frac{g^2 E}{2\pi^2 }\left\{ 1 - 4\alpha_1 \frac{\gamma^3E}{M}
- 4 \alpha_2 \left(\frac{\gamma^2E}{M}\right)^2 
+ 32 \alpha_1^2 \left(\frac{\gamma^3E}{M}\right)^2
+...\right\}
\ea
\end{widetext}
We conclude that provided that all the terms retained in the sum are small, the
transition rate of an inertial observer is insensitive to the deviations 
from Lorentz invariance.
From the first two orders we get the restrictions
 $8\alpha_1 \frac{\gamma^3 E}{M} \ll 1$ and 
 $24\alpha_2 \frac{\gamma^4 E^2}{M^2} \ll 1$,
and provided all the $\alpha_n$ are of the same order, the first condition 
is also valid for all the odd order and the second for all the even orders. 
This means that the boost factor is limited to $\gamma \ll (M/E)^{1/2}$ 
or $\gamma \ll (M/E)^{1/3}$ (depending on the presence or not of odd terms). 
Either way, this is not constraining in practice.
 
\begin{widetext}
Matters are different for UA trajectories because 
$w_n$ are time dependent.
Let us take $\alpha_1 = 0$ to simplify the calculations.
Since the Wightman function is not stationary, we remind that we calculate 
the mean rate (\ref{barR}). 
We are now integrating 
\begin{eqnarray} \label{rateTaylor}
 \bar R_\pm(\t)  &=& \frac{-g^2a}{8\pi^2}\, \re\int_{-\infty}^{+\infty}\!\!dx \, 
   e^{\mp i (2E/a) x}\frac{1}{\sinh^2\left(x-i\e\right)}
   \left(1 + \frac{\alpha_2 a^2}{4 M^2} \, 
   \frac{24 \cosh^4(a\tau - x) - 8\cosh^2(a\tau - x)-1}{\sinh^2\left(x-i\e\right)} 
    + ...\right) \qquad
\, , \nonumber \\
     &=& R_\pm^{LI} \left\{ 1 + \alpha_2 \Gamma^4(\tau) \left[ 
   4 \left(\frac{E}{M}\right)^2 + 52 \left(\frac{a}{M}\right)^2 \right]\, 
    \left( 1 + O\left( \Gamma^{-2} \right) \right)
   + ... \right\} \, .
\end{eqnarray}
\end{widetext}
where the Lorentz factor is $\Gamma = \cosh(a\tau)$.
The correction are small for times 
\ba
  \alpha_2 \left(\Gamma^2(\tau) \frac{{\rm{max}}(E,a) }{M} \right)^2 \ll 1 \, .
\ea
Since $E/M$ and $a/M$ cannot be larger than $10^{100}$ in practice, this means
\ba \label{upper bnd}
   a\tau \ll 100 \, ,
\ea
which is still compatible with the condition
\ba \label{lower bnd}
    a\tau \gg 2\pi \, ,
\ea
for the detector to thermalize with a thermal bath at temperature $a/2\pi$
(the typical circular frequency of the particles of the bath).
We can conlude that the transition rates $R_\pm$ are thermal at best within the 
interval defined by (\ref{upper bnd}) and (\ref{lower bnd}), which represents 
a few thermal periodes only (but increasing $E$ or $a$ lowers the upper bound). 
Before this, the detector has not thermalized, and beyond it, stationarity is lost.

Our final comment is on \cite{Rinaldi} in which the 
transition rates were calculated with the help of such a Taylor expansion.  
The author claims to have proved the robustness of the Unruh effect. 
He did not. 
We agree with his results, eqs. ($25$) and ($31$) (in respectively 
two and four dimensions), obtained from the expression (\ref{Rpm})
as a starting point. As the author noticed, his eqs. ($25$) and ($31$) are 
valid only for times $\tau \leq a^{-1}$.
The author failed however to recognize that 
the expression (\ref{Rpm}) is valid precisely in the opposite limit $a\tau \gg 1$.
His calculations therefore do not prove anything about the Unruh effect.

\section{Steepest descent approximation of eq. (\ref{steepest})}
\label{app:saddle}

The saddle points solutions of $f'(x_{\pm})=0$ are given in terms of 
$x_{\pm} = \ln(y_{\pm})$ with 
\ba
  y_{\pm} &=& \frac{-\eta \pm \sqrt{\Delta}}{\omega - k_z} \, , \nonumber \\
  \Delta &=& 1 + k_z^2 - \omega^2 \, .
\ea
At these points the function $f$ and its derivatives take the values
\ba\label{fff}
   f(x_{\pm}) &=& i\left( \eta x_{\pm} \pm  \sqrt{\Delta}\right) \, , \nonumber \\
   f''(x_{\pm}) &=& \pm i \sqrt{\Delta} \, , \quad f'''(x_{\pm})= - i\eta
  \, .
\ea
The saddle point approximation is good provided $\vert \Delta \vert \gg 1$, that is
$\omega \gg k_z$ and $\omega \ll k_z$.

We begin with the description of the 
positions of the saddle points in the plane $(\omega,k_z)$.
The latter is divided into three regions:
I for $0\leq \omega \leq k_z$,
II for $k_z^2 \leq \omega^2 \leq k_z^2 + 1$,
and III for $\omega^2 \geq k_z^2 + 1$.
Keeping $k_z$ fixed and increasing $\omega$ from $0$ to infinity,
the saddle points migrate as follows.

In I, $\Delta>0$ and $y^\pm\lessgtr 0$, hence 
\ba
   x_\pm = \ln\left(\frac{\sqrt{\Delta}\mp\eta}{k_z-\omega}\right) + {}^{i\pi}_{0}
\ea
The real part of $(x_+,\eta=-1)$ and $(x_-,\eta=+1)$
increases from
$\ln\left[\left( \sqrt{k_z^2 + 1} +1 \right)/k_z\right]$ to $+\infty$,
and the real part of the other two 
increases from $-\ln\left[\left( \sqrt{k_z^2 + 1} +1 \right)/k_z\right]$ to $\ln k_z$.

In II, $\Delta>0$ and $y_\pm$ is of the sign of $-\eta$, so we have
\ba
   \eta=+1 &,& x_\pm = \ln\left(\frac{1\mp\sqrt{\Delta}}{\omega-k_z}\right) + {i\pi} \\
   \eta=-1 &,& x_\pm = \ln\left(\frac{1\pm\sqrt{\Delta}}{\omega-k_z}\right) \, .
\ea
The real part of $(x_+,\eta=-1)$ and $(x_-,\eta=-1)$ 
decreases from
$+\infty$ to $\ln\left[\left( \sqrt{k_z^2 + 1} +k_z \right)/k_z\right]$,
and the real part the other two
increases from $\ln k_z$ to $\ln\left[\left( \sqrt{k_z^2 + 1} +k_z \right)/k_z\right]$.

Finally in III, $\Delta<0$ and the saddle points are complex  
\ba
  x_\pm = \ln\sqrt{\frac{\omega+k_z}{\omega-k_z}} \mp \eta i  \arctan\sqrt{-\Delta}
\ea
For each process $\eta$, both saddle points start at $\ln(k_z+\sqrt{1+k_z^2})$, one
migrates to $i\pi/2$ and the other to $-i\pi/2$.

We now wish to know which of the saddle points are actif for a given value of 
$(\omega, k_z)$. The steepest descent path $\gamma(\lambda)$, which passes through 
the actif saddle point(s), is by definition a path of stationary phase, i.e.
\ba \label{stat phase}
  \im(f[\gamma(\lambda)]) = \im(f(x_\pm)) \, ,
\ea
(or equivalently such that $\re(f(x))$ is maximum on $\gamma$),
such that the rate at which $\re(f(x))$ decreases away from the saddle point
is as high as possible, 
that is in a neighbourhood of the saddle point one requires
\ba\label{conditionfseconde}
  \vartheta \equiv \arg\left(f''(x_\pm)(x\vert_{\gamma}-x_\pm)^2 \right) = \pi \, .
\ea
The steepest descent path is a solution of both (\ref{stat phase}) and (\ref{conditionfseconde}). 
Letting $f''(x_\pm) = \vert f_\pm'' \vert e^{i2\alpha}$
and taking the origin of the parameter $\lambda$ at the saddle point, 
$x-x_\pm = \lambda e^{i\varphi_{\rm sdp}} + O(\lambda^2)$, the angle of the tangent 
of $\gamma$ is given by
\ba \label{tangent}
  \varphi_{\rm sdp} = \frac{\pi}{2} - \alpha \quad {\rm mod}(\pi) \, ,
\ea
at the saddle point.
In regions I and II, $\alpha=\pm \frac{\pi}{4}$ (for $x_\pm$ respectively, 
see eq. (\ref{fff})), hence if the steepest descent path exists, its tangent at the 
saddle point is
$\varphi_{\rm sdp} = \pm \frac{\pi}{4}$. 
In region III, $\alpha = \frac{\pi}{2}$ and
$\varphi_{\rm sdp} = 0$.

We give the curves solutions of (\ref{stat phase}) in
the parametric form $\gamma(\lambda) = \lambda + iy(\lambda)$ 
and the corresponding angles $\vartheta$ 
\ba
  \label{solI} 
   I) &&  \gamma = \lambda + i \arccos\left(
      \frac{\eta\ln\frac{\sqrt{\Delta}\mp\eta}{k_z-\omega} \pm \sqrt{\Delta}-\eta \lambda}{\omega\sinh \lambda - k_z\cosh \lambda}\right) \qquad \\
   && \lambda \to \re(x_\pm)^-  \ , \ \vartheta_\pm = \pm \pi \, \,\, ;
\nonumber \\
   &&   \lambda \to \re(x_\pm)^+ \ , \ \vartheta_\pm = 0 \\
  \label{solII}   
   II) && \gamma = \lambda + i \arccos\left(
      \frac{\ln\frac{1\mp\eta\sqrt{\Delta}}{\omega-k_z} \pm 
      \sqrt{\Delta}-\eta \lambda}{\omega\sinh \lambda - k_z\cosh \lambda}\right) \\
   && \lambda \to \re(x_\pm)^-  \ , \ \vartheta_\pm = \pi (\eta \pm 1 )/2  \,\,\, ; 
\nonumber \\  
   &&   \lambda \to \re(x_\pm)^+ \ , \ \vartheta_\pm = \pi (\pm 1 -\eta)/2 \\
  \label{solIII} 
   III) && \gamma = \lambda  \pm i\arccos\left(
      \frac{\eta\left(\ln\sqrt\frac{\omega+k_z}{\omega-k_z}-\lambda\right)}{\omega\sinh \lambda - k_z\cosh \lambda}\right) \\
   && \vartheta_\pm = \pi (1\pm 1)/2
\ea
Along the curves of constant phase (\ref{solI}) and (\ref{solII}) 
in resp. regions I and II, $\vartheta$ changes by $\pi$ as the saddle point is
passed, so the steepest descent path does not exist. 
Fig. \ref{fig12} shows $\re\left(f(x)\right)$ in the complex $x$-plane.
Indeed, the angle of the tangent of the curves is not continuous at the saddle point,
so it does not verify (\ref{tangent}).
In region III, the steepest descent path exists.
It is the curve of stationary phase (\ref{solIII}) passing through the saddle point $x_+$.

\end{appendix}

\end{document}